\documentclass[12pt,a4paper]{article} 
\pdfoutput=1
 \usepackage{jheppub}
\usepackage{enumerate}
\usepackage{epsfig} 
\usepackage{wasysym} 
\usepackage{mathrsfs} 
\usepackage{graphicx} 
\usepackage{amsfonts} 
\usepackage{amsbsy} 
\usepackage{amscd} 
\usepackage{pstricks} 
\usepackage{multirow} 
\usepackage{tikz}
\usetikzlibrary{arrows,positioning,shapes.geometric}
\usepackage{color}
\definecolor{myred}{rgb}{0.6,0,0} 
\definecolor{myblue}{rgb}{0,0.2,0.4}
\definecolor{mygreen}{rgb}{0,0.9,0.1}
\definecolor{hc}{rgb}{.9,0.1,0.7}
\definecolor{hcout}{rgb}{.9,0.7,0.9}
\definecolor{Orange}{rgb}{1.,0.65,0.}

\numberwithin{equation}{section}
\numberwithin{figure}{section}
\numberwithin{table}{section}

 \newcommand{\f}{\mathcal{F}}
\newcommand{\n}{\mathcal{N}}

\newcommand{\met}{\ensuremath{\not\!\!E_T}\xspace}
\newcommand{\be}{\begin{equation}}
\newcommand{\ee}{\end{equation}}
\newcommand{\bea}{\begin{eqnarray}}
\newcommand{\eea}{\end{eqnarray}}

\newcommand{\newc}{\newcommand}
\newc{\bi}{\begin{itemize}}
\newc{\ei}{\end{itemize}}

\newc{\ra}{\rightarrow}
\newc{\sq}   {\mbox{$\wt{q}$}}
\newc{\msq}  {\mbox{$m_{\sq}$}}
\newc{\gl}   {\mbox{$\wt{g}$}}
\newc{\mgl}  {\mbox{$m_{\gl}$}}
\def \met  {\mbox{${E\!\!\!\!/_T}$}}
\newc{\wt}{\widetilde}

\def \lspone{\wt\chi_1^0}
\def \mlspone{m_{\lspone}}
\def \lsptwo{\wt\chi_2^0}
\def \mlsptwo{m_{\lsptwo}}
\newc{\ifb}{\mbox{${\rm fb}^{-1}$}}

\def \chonepm{\wt\chi_1^\pm}

\def \mchonepm{m_{\chonepm}}

\newc{\del}{\delta}
\def \stauone{\wt\tau_1}
\def \mstauone{m_{\stauone}}
\def \slepl{\wt{l}_L}
\def \mslepl{m_{\slepl}}
\def \slepr{\wt{l}_R}
\def \mslepr{m_{\slepr}}
\def \slep{\wt{l}}
\def \mslep{m_{\slep}}
\def \lstop{\wt {t}_1}
\def \mlstop{m_{\lstop}}

\title{\begin{center}Non-universal Gaugino mass models\\ under the lamppost of muon (g-2)\end{center}}

\author[a]{Joydeep Chakrabortty,}  
 \author[b]{Arghya Choudhury,}
\author[b]{Subhadeep Mondal} 
  \affiliation[a]{Department of Physics, Indian Institute of Technology, Kanpur-208016, India} 
  \affiliation[b]{ Regional Centre for Accelerator-based Particle Physics, \\
Harish-Chandra Research Institute, Allahabad 211019, India}
\emailAdd{joydeep@iitk.ac.in}
\emailAdd{arghyachoudhury@hri.res.in}
\emailAdd{subhadeepmondal@hri.res.in}

\abstract{
In unified $\mathcal{N}=1$ supergravity scenario the gaugino masses can be non-universal. 
The patterns of these non-universalities are dictated by the vacuum expectation values of non-singlet chiral super-fields
in visible sector. Here, we have analysed the model independent correlations among the gaugino masses with an aim to explain 
the $[1\div 3]\sigma$ excess  of muon (g-2)  ($\Delta a_\mu$). We have also encapsulated the interconnections among other 
low and high scale parameters, compatible with the collider constraints, Higgs mass, relic density and flavour data. We have noted 
that the existing non-universal models are not capable enough to explain $\Delta a_\mu$ at $[1\div 2]\sigma$ level. In the process, we have also shown the impact of recent limits from the searches for disappearing track and long lived charged particles at the LHC. 
These are the most stringent limits so far  ruling out a large parameter space allowed by other constraints. 
We have also performed  model guided analysis where gaugino masses are linear combination of contributions coming from singlet and 
non-singlet chiral super-fields. Here, a new mixing parameter has been introduced. Following the earlier methodology, 
we have been able to constrain this mixing parameter and pin down the promising models on this notion. 
}

\preprint{HRI-RECAPP-2015-005}


\begin{document}
\maketitle

\section{Introduction}

Supersymmetry (SUSY) is one of the most promising beyond Standard Model (BSM) scenarios that solves the gauge hierarchy problem, 
stabilizes the Higgs mass and also addresses some of the other shortcomings of the SM. It also provides a weakly interacting massive 
particle (WIMP) which can be a viable cold dark matter (DM).  
In R-parity conserving scenario, the lightest supersymmetric particle (LSP) happens to be that DM candidate. From unification view point, 
SUSY shows an improvement over the Standard Model (SM) predictions with a consistent grand unification (GUT) scale. 

In global minimal supersymmetric standard model (MSSM), the gauge group is  same as that for the SM, but the particle content 
is extended in form of the supersymmetric partners. In unbroken SUSY scenario, the 
supertrace\footnote{Supertrace in supersymmetric theory is defined as: $\mathcal{S}tr=\sum_s (-1)^{2s} 
(2s+1) Tr(m_s^2)$, where $s,m_s$ are the spin and mass of the particle respectively.} vanishes exactly 
which is due to the degeneracy in the spectrum and equality in fermionic and bosonic degrees of freedoms in theory. 
However,  so far experimental data suggests that all the SUSY particles must 
be heavier than their respective SM partners. Thus the supertrace must be non-vanishing and, in general, proportional 
to the SUSY scale. In other words, we can say that SUSY can be realized 
in nature only in broken form\footnote{In passing we would like to draw attention to another view as 
suggested \cite{Witten:1994cga} questioning the necessity of supersymmetry breaking.}. So, the important 
question that rises in this context is following:  what is the SUSY breaking mechanism?  So far we 
have failed to pin down any specific breaking scenario, rather fenced by different possibilities. 
These mechanisms can be broadly classified into two categories: spontaneous\footnote{Dynamical 
supersymmetry breaking is an interesting possibility \cite{Witten:1981nf} but is troublesome as 
it might lead to charge and colour breaking vacua.} and explicit breaking by addition of the SUSY 
breaking soft terms in the Lagrangian \cite{Witten:1981my, Witten:1982df, Dimopoulos:1981zb,  
Nelson:1993nf, Intriligator:2006dd, Intriligator:2007py, Intriligator:2007cp}. The later one 
is not forbidden by any physical principle but it leads to enormous number of free parameters (more than 100) 
in the theory and spoils the beauty of it. We are not elaborating this possibility as this is not 
the prime moto of this paper. The outcome of spontaneous breaking depends on which field is getting 
vacuum expectation value (VEV). However, as we know, spontaneous breaking does not change the property 
of the action. Hence the supertrace which is a signature of the Lagrangian, is kept unmodified. This is 
against the experimental observation as mentioned before and thus ruled out. 
Again if we start with an exact SUSY 
Lagrangian, we are bound to break SUSY only spontaneously \cite{Witten:1981my, Witten:1982df}. 
This dilemma can be resolved if one breaks SUSY in the hidden sector. This information of SUSY 
breaking can be brought to the visible sector by different messengers, giving rise to different models, 
such as gravity-, gauge-, anomaly-mediated SUSY scenarios. 
For the sake of our work, let us further proceed our discussion on the gravity mediated SUSY breaking 
scenario: $\mathcal{N}=1$ SUGRA \cite{Ferrara:1976iq, Cremmer:1977tc, Cremmer:1977zt, Cremmer:1978iv, 
Cremmer:1978hn, Cremmer:1982wb, Cremmer:1982en, AlvarezGaume:1983gj, Nilles:1983ge, Lykken:1996xt, 
Hall:1983iz, Barbieri:1982eh}.  In the minimal SUGRA (mSUGRA) framework, there are only 5 free 
parameters compared to over 100 in the general MSSM case: universal scalar mass ($m_0$), universal gaugino mass ($m_{1/2}$), $\tan \beta$  
(ratio of the up and down type Higgs VEVs $=v_u/v_d$), 
tri-linear coupling ($A_0$) and sign of $\mu$ term ($sgn(\mu)$). However, it is not necessary for the  
gauginos to be degenerate  at the high scale itself. If the visible sector possesses an  unified gauge symmetry,
the gauginos may become non-degenerate through the VEV of the GUT breaking scalars \cite{Chamseddine:1982jx, Ohta:1982wn, Ellis:1982wr, Ibanez:1982ee, 
Inoue:1982pi, Ibanez:1982fr, Ellis:1985jn, Drees:1985bx, Ellis:1984bm, Chakrabortty:2010xq, 
Bhattacharya:2009wv, Martin:2009ad, Chakrabortty:2008zk}. Then the high scale input parameters 
consist of $m_0, M_3, \tan\beta, A_0$ and $sgn(\mu)$, where, $M_3$ is the mass scale for the $SU(3)_C$ gauginos. This also 
determines $M_1$ and $M_2$, as they are correlated: $M_1:M_2:M_3=a:b:c$, where $a,b,c$ depend on patterns 
of the symmetry breaking. This enriches the possibility of having different correlations among 
these gaugino masses which lead to different compositions of the LSP at low scale. Thus unlike 
the mSUGRA case, here, the LSP can be bino, wino, higgsino or admixture of these three states. This allows us to 
 explore wide range of phenomenologies driven by non-universal gaugino masses. 
The phenomenological aspects of non-universal gaugino mass scenario are discussed in \cite{Li:2010xr, 
Atkins:2010re, Balazs:2010ha, Wang:2015mea, Chakraborti:2014fha, Miller:2014jza,  Martin:2013aha, 
Chakrabortty:2013voa,  Bhattacharya:2013uea, Miller:2013jra, Younkin:2012ui, King:2007vh, Corsetti:2000yq, 
Chattopadhyay:2001mj, Anderson:1999uia, Huitu:1999vx, Huitu:2008sa, Chattopadhyay:2009fr, Das:2014kwa,
Kaminska:2013mya, Bhattacharya:2007dr, Abe:2007kf, Khalil:1999ku, Chakrabarti:2004ps, 
BirkedalHansen:2002am, Cerdeno:2004zj, Bityukov:1999am, Bityukov:1999sb, Bityukov:2001yf, 
Bityukov:2002fs}. Due to its varieties of LSP configurations, these models have significant 
impact on the analyses regarding DM searches and muon (g-2) excess. These two issues  
have been  explored extensively within mSUGRA framework. But  the present limits on squarks 
and gluino put severe stringent bounds on mSUGRA parameter space. 
This results in unavailability of SUSY spectrum within this framework that can explain 
$\Delta a_\mu$\footnote{$\Delta a_{\mu}$ denotes the discrepancy between experimentally measured 
value of muon (g-2) and the SM predicted one. 
} at $[1\div 2]\sigma$ level respecting flavour constraints.

This has been the motivation of our work. The prime aim of our  analysis is to understand the model 
independent correlations among the MSSM gauginos which can successfully explain the anomalous muon (g-2) 
excess over the SM, and simultaneously satisfy different  experimental bounds at low energy. In this 
context we have  discussed  two scenarios: non-universal gaugino \& universal scalars, universal 
gaugino \& non-universal scalars. We have extended our former framework by adopting existing  SUSY-GUT 
models and considering the general gaugino spectrum. We also introduce here a new mixing parameter 
that is related to the superpotentials. We have encapsulated the range of this new parameter while 
explaining muon (g-2) excess at $[1\div 3]\sigma$ level.


\section{Generation  of  Gaugino Masses in $\mathcal{N}=1$ unified SUGRA models}
\label{mass_gen}

Here, we have briefly reviewed the Lagrangian based discussion on the generation of gaugino masses.
In $\n=1$ supergravity framework, the part of the needful Lagrangian which is associated with the gauge 
and gaugino sectors is expressed as \cite{Cremmer:1978iv, Cremmer:1978hn, Cremmer:1982wb, Cremmer:1982en, 
Ellis:1982wr, Ibanez:1982ee, Inoue:1982pi, Ibanez:1982fr, Ellis:1985jn, Drees:1985bx, Ellis:1984bm} 
\begin{eqnarray} \label{lagrangian_sugra_gauge}
{\bf e}^{-1} \mathcal{L} &=& -\frac{\f_{\alpha \beta}}{2} \frac{\overline{\lambda^\alpha} \slash \! \!\!\! D \lambda^\beta}{2}
 - \frac{1}{4} Tr[F_{\mu \nu} \f(\Phi) F^{\mu \nu}]  \nonumber \\
         & + & \frac{1}{4} e^{-G/2} G^i (G^{-1})^j_i \left[\frac{ \delta \f^{*}_{\alpha \beta}(\Phi^{*})} {\delta \Phi^{j*}} \right] \lambda^{\alpha} \lambda^{\beta} + h.c.,            
\end{eqnarray}
where, $\alpha,\beta,\gamma =[1,..., Adj]$\footnote{For $SU(N)$ and $SO(N)$, $Adj$ is defined as $(N^2-1)$ and $(N^2-N)/2$ respectively.}.
Here, we have set $M_{Pl}/\sqrt{8\pi}=1$.  Our nomenclature is following: $G,\f, \Phi$ and $\lambda$ are the 
K$\ddot{a}$hler potential, the super-potential, chiral super-fields and gaugino fields respectively.

\begin{figure}[h!]
\centering
\hspace{-1cm}
\includegraphics[scale=0.62,angle=0]{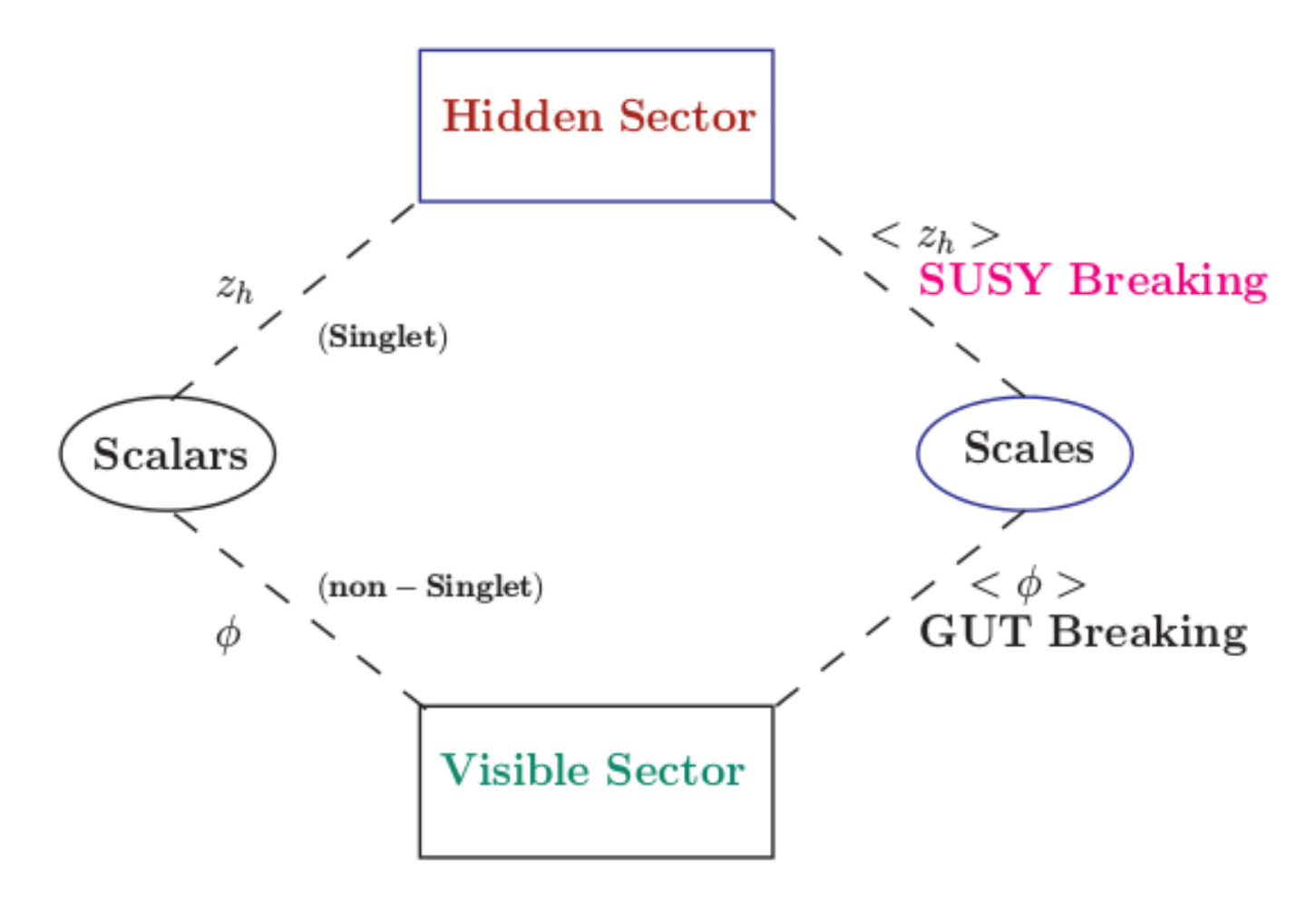}
\includegraphics[scale=0.21,angle=0]{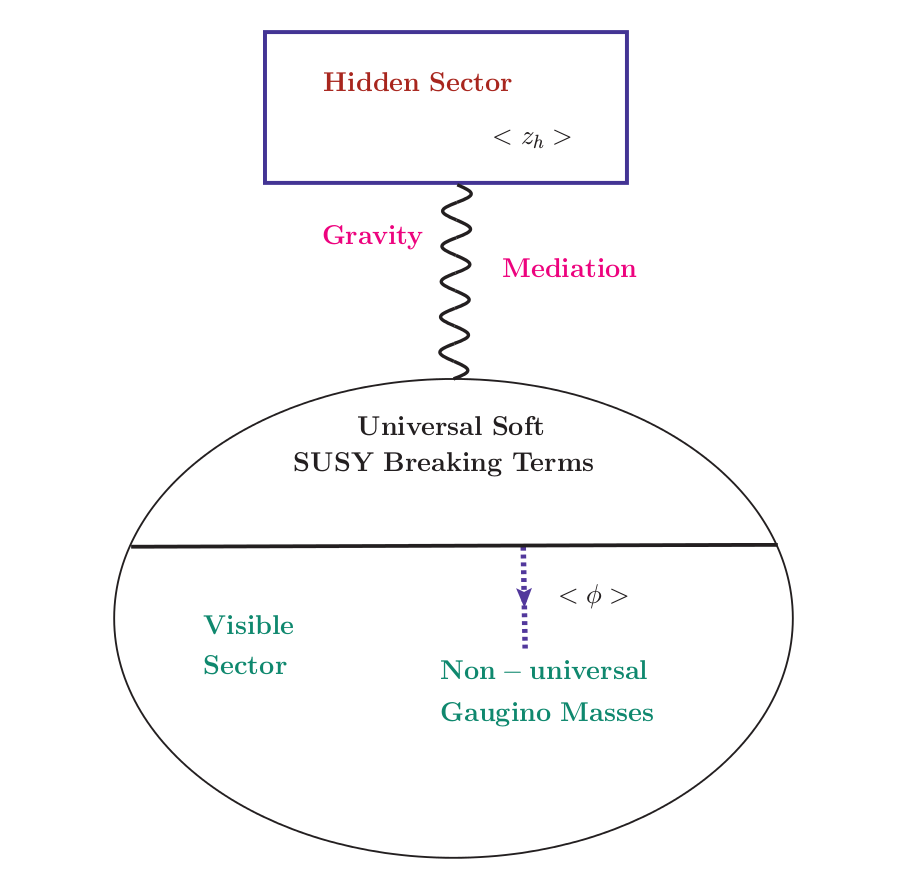}

\caption{Schematic flow chart to demonstrate spontaneous breaking of supersymmetry in the hidden sector 
through the VEV of the singlet chiral super-field ($z_h$) and successive breaking of GUT symmetry in the 
visible sector through the VEV of non-singlet chiral super-filed ($\phi$).  As the outcome of  this process, the soft SUSY
 breaking terms are generated in the visible sector and the gauginos become non-degenerate at the high scale itself.}
\label{fig:SUGRA_Gaugino_Mass}
\end{figure}

Here, the derivatives of K$\ddot{a}$hler potential are defined as:  
$G^i = \frac{\delta G}{\delta \Phi_i}, G^i_j= \frac{\delta^2 G}{\delta \Phi^{*j} \delta \Phi_i}$, and $(G^{-1})^i_j$ is the inverse of $G^i_j$.
The chiral function $\f_{\alpha \beta}(\Phi)$ is an analytic function of $\Phi$.
These $\Phi$'s are the chiral super-fields belonging to the visible $(\phi)$ as well as the hidden $(z_h)$ sectors. 
The chiral field $z_h$ is singlet under the gauge symmetry of the visible sector.  The other chiral super-field $\phi$ is non-singlet 
under such gauge symmetries. The choices of $\phi$ are further restricted from the gauge kinetic term $Tr[F_{\mu \nu} \f(\Phi) F^{\mu \nu}]$ 
which dictates that $\phi$ can only belong to the symmetric product of two adjoint representations. 
In this framework, both SUSY and unified gauge symmetry are broken spontaneously. The VEV of $z_h$ ($z_h^0$) that breaks local 
SUSY spontaneously is very large ($\sim 10^{19}$ GeV) where the VEV of $\phi$ $(V_{\rm GUT})$ sets the GUT scale ($M_{U}$). 
The generic structure of the chiral function, $\f_{\alpha \beta}$, can be given as suggested in \cite{Ellis:1984bm, Ellis:1985jn} 
(neglecting higher order terms which are more suppressed):
\begin{equation} \label{eq:superpotential}
\f_{\alpha \beta} = \f_1 ~\delta_{\alpha \beta} + \f_{2}~ c_{\alpha \beta \gamma} ~\phi^\gamma,
\end{equation}
where $\f_{1,2}$ are the functions of chiral super-fields of hidden and visible sectors, and $c_{\alpha \beta \gamma}$ are the group theoretic factors. The first term is 
the canonical term leading to mSUGRA like scenario where all the gauginos are degenerate, i.e., universal. 
The second term is due to the presence of GUT symmetry breaking scalars which cause splitting in universal 
gaugino spectrum of mSUGRA leading to non-universal gaugino masses at the high scale itself. The different 
choices of these $\phi$ fields lead to  possible non-universalities as the high scale boundary 
conditions. For the purpose of our present analysis, we consider that 
$\f_{1,2}$ are the only functions of $z_h$, and $\phi$. 
In Fig.~\ref{fig:SUGRA_Gaugino_Mass}, we have summarized this whole process through a schematic flow chart.  

Once SUSY and GUT are broken, one can recast the gaugino mass terms as
\begin{equation}\label{eq:nonug_gaugino_mass}
M_i  = M ~\left[P~ +~ \delta_i~  V_{\rm GUT}~ Q\right],
\end{equation}
where the $\delta_i$'s are the  group theoretical factors for broken and unbroken generators 
(see Refs.~\cite{Ellis:1984bm, Ellis:1985jn, Drees:1985bx} for details) similar to 
$c_{\alpha \beta \gamma}$ in Eq.~\ref{eq:superpotential}. We can 
write down $i \ni [j,X]$ where $j$ and $X$ represent the indices for unbroken and broken generators 
respectively. To understand this better, let us consider the breaking of  
$SU(5)$ to $U(1)_Y\otimes SU(2)_L \otimes SU(3)_C$ ($\mathcal{G}_{MSSM}$).   
As an outcome of this mechanism, 12 generators of $\mathcal{G}_{MSSM}$ are unbroken, and rest of the 
$(24-12)=12$ generators are broken. In our further analysis, we denote the gauginos corresponding 
to the broken generators as $X$ and others are  associated with index $j$. Here, $P$ and $Q$ are the functions 
of $z_h^0$, $V_{\rm GUT}$ and derivatives of $\f_{1,2}$, evaluated at  [$z_h^0$, $V_{\rm GUT}$]. Within 
SUGRA framework, unfortunately, the exact functional forms of  $\f_1,\f_2$ are unknown and same for  $P,Q$. 
However, in simplified scenario, like mSUGRA, due to  absence of GUT  group
in the visible sector there is no need of  non-singlet field $\phi$ in the theory. This implies that all the 
$\delta_i$'s are identically zero leading to all the gaugino masses to be degenerate, i.e., universal.

But within the unified SUGRA framework, the generic form of the gaugino mass is given in 
Eq.~\ref{eq:nonug_gaugino_mass} which cannot be further simplified to understand the specific 
correlations among them unless we make some assumption.  
In Refs.~\cite{Chakrabortty:2010xq, Bhattacharya:2009wv, Martin:2009ad, Chakrabortty:2008zk}   
it was pointed out from pure  phenomenological perspective that if $P/Q$ is negligibly small 
then the ratio of the gaugino masses are just the ratios of $\delta_i$'s. 
It has been also argued in Refs.~\cite{Ellis:1984bm, Ellis:1985jn} that to stabilise 
the cosmological constant, the masses of the $X$-gauginos must vanish and also provide unique 
mass ratios for MSSM gauginos independent of $P$ and $Q$. Here, the gaugino masses are automatically 
non-universal and posses definite non-universal patterns. To complete this discussion, we would 
like to mention that the generation of gaugino masses in context of supergravity framework was 
first discussed in Refs.~\cite{Ellis:1984bm, Ellis:1985jn, Drees:1985bx} for $SU(5)$ unified theory. 
Later this has been generalised for   $SO(10)$ and $E(6)$ GUT groups in 
Refs.~\cite{Chakrabortty:2010xq, Bhattacharya:2009wv, Martin:2009ad, Chakrabortty:2008zk}.

The generic form of gaugino masses, without any further assumption,  can be recast as:
\begin{equation}\label{eq:nonug_gaugino_mass_general}
\mathcal{M}_i  = M^{'} ~\left[1 +~\wp~ \delta_i~ \right],
\end{equation}
where $\wp$ is the ratio of $P,Q$ and can be thought of as a measure of mixing between singlet 
and non-singlet contributions. We will emphasize  more on this mixing parameter, $\wp$,
from muon (g-2) point of view for some specific models.


\section{Muon g-2}
The precision measurements are leaving very little room for new physics.  
One of them which has been measured experimentally with an immaculate accuracy is 
the muon anomalous magnetic moment. In the SM, there exists a tree level contribution 
to muon($\ell_\mu$)-muon-photon($A_\rho$) coupling  ($ie\overline{\ell_\mu} \gamma^\rho \ell_\mu  A_\rho$). 
Along with that in BSM scenario, the relatively heavier and so far unobserved  particles may contribute to 
this vertex through radiative corrections in the form of an effective operator like 
$(ie/4m_\mu) a_\mu \overline{\ell_\mu} [\gamma^\lambda , \gamma^\rho] \ell_\mu 
(\partial_\lambda A_\rho - \partial_\rho A_\lambda)$.
So far, there is a discrepancy, $\Delta a_\mu$, among the SM theoretical prediction and experimentally 
observed value  \cite{Bennett:2006fi, Roberts:2010cj}. Thus the BSM contributions, if there is any, 
have to be fitted within the deviation \cite{Nyffeler:2013lia} 
\begin{equation}
\Delta a_\mu = (29.3 \pm 9.0 ) \times 10^{-10}.
\end{equation}

\begin{figure}[ht]
\centering
\includegraphics[scale=0.90,angle=0]{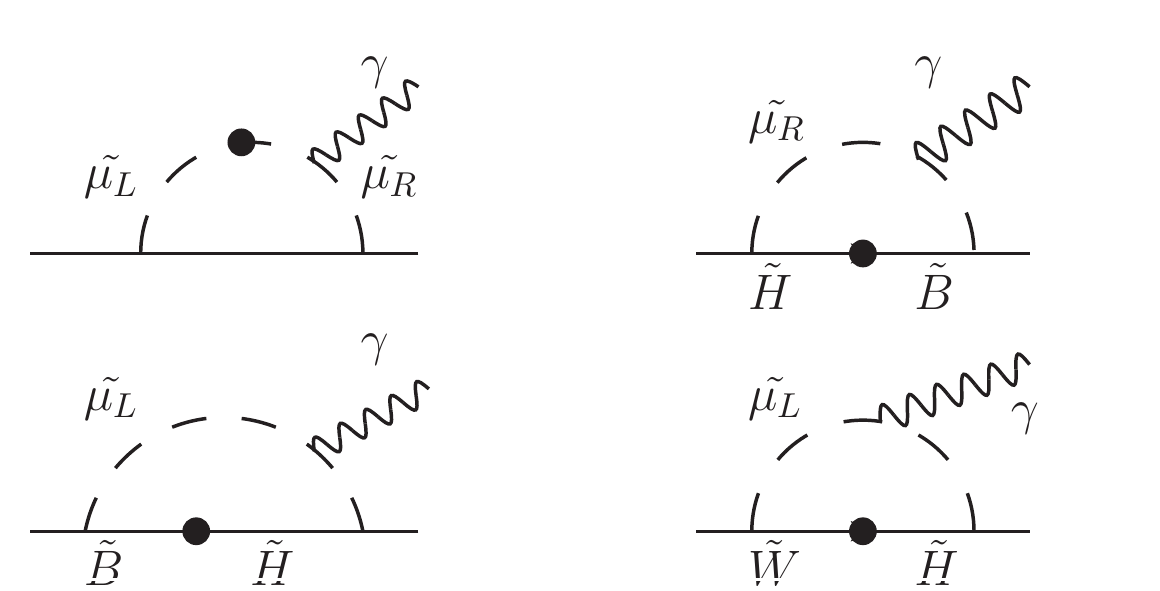}
\caption{Feynman diagrams that contribute to muon (g-2) involving neutralinos ($\tilde{B},\tilde{W},\tilde{H}$) and smuons $(\tilde{\mu}_{L,R})$.}
\label{fig:muong-2_NSM}
\end{figure}

\begin{figure}[ht]
\centering
\includegraphics[scale=0.90,angle=0]{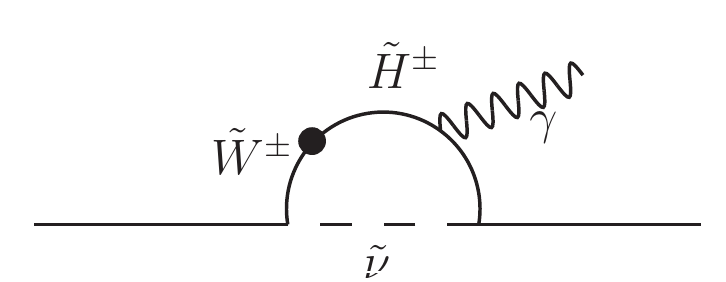}
\caption{Feynman diagrams that contribute to muon (g-2) involving charginos ($\tilde{W}^{\pm},\tilde{H}^{\pm}$) and sneutrino $(\tilde{\nu})$.}
\label{fig:muong-2_CSN}
\end{figure}

It is expected that SUSY can explain this excess through the exchanges of different sparticles 
and that has been one of the motivation for TeV scale SUSY. The loop induced contributions (see Figs.~\ref{fig:muong-2_NSM}, \ref{fig:muong-2_CSN}) 
involving sleptons, neutralinos, charginos and sneutrinos are very important in this  regard \cite{Kosower:1983yw, Yuan:1984ww, Fargnoli:2013zia, Fargnoli:2013zda}. 
In the context of general SUSY 
scenario, muon (g-2) has been discussed in \cite{Moroi:1995yh, Heinemeyer:2003dq, Stockinger:2006zn, Cho:2011rk, Akula:2013ioa, Endo:2013lva,  Endo:2013bba, 
Chakraborti:2014gea}.  Within the unified SUGRA framework the contributions to anomalous muon magnetic moment are discussed in 
\cite{Lopez:1993vi, Chattopadhyay:1995ae,  Chattopadhyay:2001vx, Babu:2014lwa, Mohanty:2013soa, Chakraborti:2014gea, Gogoladze:2014cha, Ajaib:2015ika}.

In this paper, we have started with a framework where all the scalar masses are universal but the 
gauginos are non-degenerate at the high scale itself. The other parameters, say, $\tan\beta,~ 
sgn(\mu),~ A_0(=-2m_0), ~m_{H_u}^2=m_{H_d}^2(=m_0^2)$ are chosen suitably. Our intention is to adjudge the 
patterns of non-universalities in  gaugino masses under the light of $\Delta a_\mu$ at $[1\div 3]\sigma$ level. 
In other words, we have used $\Delta a_\mu$ to intrigue unified SUGRA model building. 

Computation of the radiative contributions to $a_\mu$ is very similar to the generic SUSY 
scenario. The only difference is in the generation of the low scale SUSY spectrum. Due to the high scale 
boundary conditions, our scenario is much more constrained. Thus the low scale  sparticle spectrum 
cannot be tuned arbitrarily to fulfil  the necessary contributions for muon (g-2). In general the 
chargino-sneutrino loop (see Fig.~\ref{fig:muong-2_CSN}) contributes more dominantly compared to the 
neutralino-smuon loop (see Fig.~\ref{fig:muong-2_NSM}). But this is not always true as these 
contributions depend on the right-handed smuon $(\tilde{\mu}_R)$ mass.

The mas  parameters masses of sleptons, neutralinos, charginos and sneutrinos  play crucial roles to determine the muon anomalous magnetic moment. The high scale parameters which are involved in these computations 
are $M_{1,2},  m_0(\equiv \tilde{\mu}_{L,R}), \tan \beta, A_0$ and $sgn(\mu)$. 
The dependence of $\Delta a_\mu$ 
on $\tan \beta$ is linear which is due to the requirement of chirality flipping through the Yukawa couplings. 
Thus $\Delta a_\mu$ is very sensitive to $\tan \beta$  and larger value of  
$\tan \beta$ is favoured to produce bigger contribution to $\Delta a_\mu$. 
However, we are forced to respect the constraints from flavour data and need to be careful 
while considering large  $\tan \beta $. We would like to notify that the choice of $sgn(\mu)$ 
is also restricted as $\mu M_3<0$ is severely constrained 
from the measurement of BR($b \to s \gamma$). Also note that $\Delta a_\mu$  prefers $sgn(\mu)>0$. 
Hence throughout this work, we only consider $\mu>0$.

In our analysis, the scalar masses are universal at the high scale, thus the splitting between left- and 
right-handed sleptons are not large. However, through the renormalisation group evolutions, the off-diagonal 
terms in the slepton mass matrix can be generated and that may lead to an open possibility of having 
contributions from all generations of sleptons. Again if the $\mu$ term, i.e., the higgsino mass 
parameter is larger than the masses of the left-handed smuons then the masses of the right-handed 
smuons play crucial role (see left-top of Fig.~\ref{fig:muong-2_NSM}). Then the contribution to $\Delta a_\mu$ 
from this diagram decreases as mass of the right-handed smuon increases. Note that this effect is only visible if 
the diagrams in Fig.~\ref{fig:muong-2_NSM} dominate over the one in Fig.~\ref{fig:muong-2_CSN}. In general, if the lightest 
neutralino, lighter chargino,  left- \& right-handed smuons and sneutrinos are nearly degenerate 
(within few tens of GeV) and also $\mu$ term is  of the same order, then the following diagrams 
dominantly contribute positively to $a_\mu$ (see top-left of Fig.~\ref{fig:muong-2_NSM} and Fig. \ref{fig:muong-2_CSN}).  
Note that if the sneutrinos are lighter than the smuons and also the $\mu$ term (as in our case), 
the dominant contribution may come from Fig.~\ref{fig:muong-2_CSN}.

\section{Collider, Low energy and Dark Matter Searches Constraints}
\label{constraints}

The recent searches at LHC continue to put severe exclusion limits on sparticle masses 
and couplings. In case of SUSY, the parameter space is very sensitive to the mass of the 
observed Higgs boson and other flavour data. Along with these collider constraints one cannot 
ignore the impact of DM searches (direct, indirect) and also needs to respect 
the upper bound of relic density.  Here, we catalogue the imposed constraints in our 
analysis.

\subsection{Collider Constraints}
In this section, we first discuss the limits obtained from LEP (I \& II) and then from the LHC searches. 
The latest search results for long-lived charged particles and disappearing tracks at the LHC 
are also crucial for our study.  Thus we prefer to discuss these separately at the end of this subsection.

\subsubsection*{LEP exclusions:}

We have implemented the following LEP constraints on the masses of sparticles: 
\bi
\item sleptons: $\mslepl, \mslepr > 100.0 $ GeV where $l = e,\mu$; $\mstauone > $ 86.6 GeV,
\item lighter chargino : $\mchonepm >$ 103.5 GeV.
\ei
More details on these exclusions are discussed in Ref.~\cite{LEPSUSYWG}.

\subsubsection*{LHC exclusions:} 

After the completion of LHC Run-I, the exclusion limits on the sparticle masses 
are much more stronger now. More specifically, due to the large production cross-sections,  
constraints on the masses of strongly interacting SUSY particles (e.g., gluino, squarks) 
are more restrictive. Both ATLAS and CMS collaborations have updated their 
analyses in $n$-leptons+$m$-jets (with or without $b$-tagging)+$\met$ (where $n$ and $m$ can take values 0, 1, 2,...) 
channels for supersymmetric models like mSUGRA, cMSSM and other simplified SUSY scenarios. 
Apart from the production cross-section, all the limits also crucially depend on the branching 
ratios (BRs) and mass separations among the mother and daughter sparticles. 

In SUGRA type scenarios, all the gaugino ($\lspone, \chonepm, \gl$) masses are correlated:
either universal or non-universal  at the GUT scale itself.  As  we are focussing on non-universal 
gaugino scenario, we have incorporated the limits coming 
from the searches of simplified models with suitable changes\footnote{The results obtained 
in simplified  models are mainly based on the assumption 
of single particle pair production and one or two step decay modes. Thus one needs to  be very 
careful while using these bounds. The ideal situation would be to calculate the production 
cross-section for each and every SUSY spectrum  and compute the  
the LHC bounds accordingly.  But this is beyond the scope of our analysis. 
The methodology that has been adapted in this paper is conservative and more importantly not 
capable enough to alter our predictions.} for the sake of our analysis rather using direct mSUGRA bounds.

We have mentioned earlier that at the electro-weak (EW) scale, the contributions 
to muon (g-2) are controlled by the following parameters: masses of sleptons, lighter chargino, 
lightest neutralino and their compositions. 
 As we are working within an unified framework at the high scale, the low scale spectrum 
is correlated with the high scale soft SUSY breaking terms and the hierarchy in low scale spectrum is 
determined by other parameters too. For example,  one can accommodate 
light sleptons even satisfying present limits on heavy gluino ($\mgl$) and squarks ($\msq$) masses. 
\\

\textbf {Higgs mass:} 
We allow only those high scale parameters that lead to the lightest Higgs boson mass to be: $M_h = (122\div 128)$ 
GeV considering a theoretical uncertainty of $\sim 3$ GeV\cite{Aad:2012tfa, Chatrchyan:2012ufa}. \\

\textbf {Limits on chargino and neutralino masses:} 
Most of the LHC analyses on direct chargino searches \cite{Aad:2014nua, Khachatryan:2014qwa}
are based on the assumption that the LSP is bino type and the lighter chargino is wino type. 
For higgsino or mixed type LSP and(or) chargino, the production cross-sections of 
$\chonepm \lsptwo$ reduce significantly and the limits from trilepton data 
\cite{Aad:2014nua, Khachatryan:2014qwa} become much weaker. 
Again the limits are slightly sensitive to the masses of sleptons\footnote{For varying slepton masses and 
different mass hierarchies, using the LHC 8 TeV data the limits are revisited in a recent analysis 
\cite{Chakraborti:2014gea}. For revised mass limits in higgsino dominated and(or) mixed 
chargino scenarios see  \cite{arghya}.} 
where the  LHC collaborations have assumed the slepton masses to be in the midway between
$\mchonepm$ and $\mlsptwo$.
For bino type LSP and wino type lighter chargino, we have implemented the conservative limits in our analysis.
Here, we have supplemented the mass limits mentioning specific hierarchical scenarios:
\bi
\item	If $\mslep < \mchonepm$, $\mlsptwo$:  for $\mlspone < $ 300 GeV, $\mchonepm$ is excluded 
	below 720 GeV \cite{Aad:2014nua, Khachatryan:2014qwa}. 
	With an conservative approach having $\mlspone$ in between 300 to 400 GeV,
	$\mchonepm$ is excluded  in the range of 400 to 700 GeV and for $\mlspone >$ 400 GeV, 
        the limits are not applicable\cite{Aad:2014nua, Khachatryan:2014qwa}. 
	
\item   For heavy slepton scenarios ($\mslep > \mchonepm$, $\mlsptwo$):
	$\chonepm$ and $\lsptwo$ decay via $W$ and $Z$ bosons and the limits 
	on chargino masses are relatively weaker. 
	Again the lighter chargino mass below 420 GeV is excluded with $\mlspone \lesssim$ 140 GeV \cite{Aad:2014nua}.  

\item  	When the $\lsptwo \ra h \lspone $ decay mode dominates, the limits become 
	much weaker in the heavy slepton scenarios. Masses of $\chonepm$ and $\lsptwo$ 
        are excluded upto  240 GeV for a massless $\lspone$ and the limits 
       are invalidated for  $\mlspone >40$ GeV \cite{Aad:2015jqa}.  
\item   With light third generation slepton ($\mstauone < \mchonepm$, $\mlsptwo$):
         $ \mchonepm$ upto 380 GeV is excluded for $\mlspone < 85$ GeV  \cite{Aad:2014nua}.  
\ei


{\bf Limits on slepton masses: } 
ATLAS and CMS have searched for charged sleptons (first two generations) from direct 
production of R(ight) and(or)  L(eft)-type sleptons \cite{Aad:2014vma, Khachatryan:2014qwa}. 
ATLAS limits are slightly stronger than that suggested by CMS. 
Though we have implemented the limits for R-, L- and R=L-types separately, 
here, we have discussed only that for R=L-type slepton scenario
as suggested by  ATLAS collaboration using di-lepton search channel \cite{Aad:2014vma}.
The masses of selectron and 
smuon (both L- and R-type) are excluded in the mass region between [$90\div325$] GeV with  massless $\mlspone$. 
But the sensitivity of exclusion limit is relaxed once the slepton-LSP mass splitting decreases. 
For example, for $\mlspone$ = 100 GeV, common left- and right-handed slepton masses within 
$[160 \div 310]$ GeV are excluded. We have adopted the conservative  bin wise limits on slepton masses from 
Fig.~8 of Ref.~\cite{Aad:2014vma}. For example, slepton mass is excluded within $[268 \div 310]$ GeV 
for $\mlspone$ [170$\div$180] GeV. 
We have to keep in mind that these limits exist for LSP mass upto 180 GeV. \\

{\bf Limits on gluino mass: }
At the LHC, squarks and gluino pair production cross-sections are the largest among other SUSY production  channels.
The limits on gluino mass ($\mgl $) crucially depend on the decay properties of $\gl$ and(or) mass hierarchy between 
squarks, gluino and other sparticles. 
If the squarks are lighter than the gluino ($\mgl > \msq$), gluino decays via $\wt q q$ and 
for such scenario when all squarks and gluinos are produced, the limits on the strong sector of 
sparticle is most stringent. In mSUGRA (cMSSM) type of scenarios with $\tan \beta$ = 30, $A_0$ = -2$m_0$ and $\mu >$ 0, 
degenerate squarks and gluinos  are excluded for masses upto 1.7 TeV at 95 $\%$ CL \cite{Aad:2014wea}.  
This limit is applicable for  relatively small universal scalar mass $m_0$ (typically $<$ 1 TeV). 
For large $m_0$ and small $m_{1/2}$ (say, $\sim [500 \div 600]$ GeV), gluino 
decays into $\lstop t$ (note that, the first two generation squarks are much heavier than stop). 
In such scenarios (with top quark dominated final states), gluino masses smaller 
than 1.4 TeV are excluded  from $[0 {\rm -} 1]\ell + 3~b~jets + \met$ analysis \cite{Aad:2014lra}.  

In simplified scenarios where gluino decays as: $\gl  \ra \wt q q \ra  q \bar q \lspone$ 
($q$ denotes  first two generation squarks) then
$\mgl$ upto $1.5$ to $1.55$ TeV is excluded for $\mlspone$ upto 600 GeV \cite{Aad:2014wea}.  
When the squarks are very heavy and gluino decays via $\gl  \ra q \bar q \lspone$ 
(first two generations) then $\mgl$ below 1.4 TeV is excluded for $\mlspone \lesssim$ 300 GeV \cite{Aad:2014wea}.
All these limits are weakened considerably for the compressed scenarios\footnote{In compressed scenarios, due to small mass difference between the mother 
and daughter particle, the missing energy ($\met$) and transverse momentum ($p_T$) of the jets or leptons 
become softer. Eventually the limits become weaker \cite{LeCompte:2011fh, Bhattacherjee:2012mz, Dreiner:2012gx, Bhattacherjee:2013wna, Mukhopadhyay:2014dsa}.}. 
For example,  when the difference between $\mgl$ and $\mlspone$ is very small then 
the exclusion limits on $\mgl$ reduces to $\sim$ $[550 \div  600]$ GeV (for details see Fig.~10a of Ref.~\cite{Aad:2014wea}). 
Even for other scenarios or decay patterns, gluino mass below $1.1$ to $1.3$ TeV is excluded 
for relatively light neutralino.  The detail reports on SUSY searches regarding various limits for different scenarios, one can consult Refs.~\cite{CMSSUSY, ATLASSUSY}.\\


{\bf Limits on squarks mass: } 
As discussed earlier, squark masses are excluded upto 1.7 TeV when they are 
degenerate  with gluino. For very heavy gluino scenarios we have implemented 
the following constraints in our analysis: 
\bi
\item   Light flavoured squark masses are excluded below 850 GeV for $\mlspone \leq$ 350 GeV \cite{Aad:2014wea}. 
\item   For the decay modes $\lstop \ra t \lspone$, the exclusion limit on $\mlstop$ is upto $[600 \div 700]$  
	GeV for  $\mlspone \leq$ 250 GeV \cite{Aad:2014kra, Aad:2014bva}. 
	Again when stop decays as $\lstop \ra b \chonepm$,  depending on 
	the assumption over the chargino masses, ATLAS and CMS collaborations have excluded 
	stop mass upto $[500 \div 600]$ GeV for $\mlspone$ upto $[200 \div 250]$ GeV \cite{Aad:2014kra,Aad:2014qaa}. 
	For other decay modes, like  $\lstop \ra c \lspone$; $b W \lspone$; $ b f f^\prime \lspone$, 
	these limits become weaker: $\mlstop \geq$ $[240 \div 260]$ GeV \cite{Aad:2014kra, Aad:2014qaa, Aad:2014nra}.  
	\item Mass of sbottom below 620 GeV is excluded at 95\% CL when LSP mass is $\leq$ 150 GeV 
	\cite{Aad:2013ija}. For small mass difference between sbottom and LSP, exclusion limit is pushed upto 250 GeV \cite{Aad:2014nra}. 
\ei 

\subsection*{ LHC searches on heavy charged particles:}

\subsubsection*{I. Limits from the search for disappearing track}\label{sec:disappearing_track}
ATLAS and CMS have recently presented searches for charginos 
based on the high-$p_T$ disappearing 
tracks\footnote{ In the nearly degenerate scenarios, 
$\chonepm$ decays via $\pi ^\pm \lspone$ \cite{Chen:1996ap}.
Due to the small mass difference between $\chonepm$ and $\lspone$, the phase space is limited 
and the chargino has a significant lifetime. On the other hand, 
the daughter pion has momentum of $\sim$ 100 MeV which is 
typically too small for its track to be reconstructed. For charginos
that decay inside the tracker volume resulting in a disappearing track.}
when the $\chonepm$ is nearly degenerate with $\lspone$.  
For $\Delta M$ = $\mchonepm -\mlspone$ = 140 (160) MeV, 
ATLAS and CMS have excluded $\mchonepm$ upto 500 (270) GeV,
see Fig.~5 in \cite{CMS:2014gxa} (CMS) and Fig.~7 in \cite{Aad:2013yna} (ATLAS) for details. 
We incorporate these exclusion contours obtained by ATLAS and CMS to find the compatible parameter 
space in our analysis. We have to keep in mind that this limit is applicable only when both the lightest neutralino 
and lighter chargino are  wino like.

\subsubsection*{II. Limits from the search for long lived charged particle}\label{sec:long_lived} 
ATLAS collaboration has  published the exclusion limits which are out come of  searches for heavy 
long-lived charged particles using 8 TeV data\footnote {When charged particles travel with speed 
slower than the speed of light, they can be identified and their mass can be 
determined from their measured speed and momentum. ATLAS collaboration 
has measured these quantities using  time of flight and specific 
ionisation energy loss.} \cite{ATLAS:2014fka}. When the charginos ($\chonepm$) are nearly degenerate with 
$\lspone$, i.e., $\Delta M  <$ 135 MeV, then the exclusion limit on chargino mass becomes more 
stringent: $\mchonepm >$ 620 GeV \cite{ATLAS:2014fka}. CMS has also presented a similar  analysis \cite{Khachatryan:2015jha, CMS-PAS} 
and the results are in well agreement with ATLAS.

We have noted that these two above mentioned searches 
play crucial roles and lead to most stringent constraints. We have discussed the impact of these new limits
in the following sections.
In passing we would like to mention that both these exclusion limits, for
nearly degenerate  scenarios, crucially depend on the 
composition of chargino. For higgsino dominated lighter chargino 
these limits are relaxed considerably.

\subsection{Flavour physics data}

\bi 
\item The measured value of BR$(b\rightarrow s\gamma)$ does agree moderately well with the SM prediction and leaves very little room to fit BSM contribution within it. 
Thus this BR turns out to be a severe  constraint whose impact can not be unnoticed.
In the MSSM framework,  the charged Higgs and chargino exchange diagrams may contribute dominantly to this branching ratio. 
Since light chargino  boosts  the required enhancement in $\Delta a_{\mu}$, this constraint is 
very important for our study. However, the contributing diagrams to BR$(b\rightarrow s\gamma)$
interfere destructively if $\mu$ and $A_t$ are of opposite signs. Hence we choose to work with a positive $\mu$ and negative $A_t$ throughout. We impose the following 
constraint for all our points: $2.77 \times 10 ^{-4} < $ BR($b \ra s \gamma$)   $ < 4.09 \times 10 ^{-4}$ (at 3$\sigma$ level) \cite{Lees:2012ym}.
\item The flavour physics constraint coming from the measurement of BR$(B_s\rightarrow \mu^+ \mu^-)$ puts strong bounds on the MSSM parameter space. 
For large $\tan \beta$, this branching ratio is proportional to 
$(\tan\beta)^6$ and inversely proportional to $m_A^4$. Thus this is expected to be a critical constraint for large $\tan\beta$ 
scenarios which are favoured to boost the enhancement in $\Delta a_{\mu}$. We impose $0.67 \times 10 ^{-9} < $ BR($B_s \ra \mu^+ \mu^-$) 
 $ < 6.22 \times 10 ^{-9}$ (at 2$\sigma$ level) \cite{Aaij:2013aka, Chatrchyan:2013bka} as a constraint throughout our study. 
\ei

\subsection{Dark Matter Constraints}
\subsection*{I. Relic density}

In this work, we have combined WMAP nine year data \cite{Hinshaw:2012aka} (2$\sigma$ bound) with 10\% 
error in theoretical estimation which together propel the upper bound of relic density to 0.138. Thus 
including this range, the 3$\sigma$ limit as suggested by the PLANCK \cite{Ade:2013zuv} can be written as: 
\begin{equation}
0.092 < \Omega h^2 < 0.138.
\end{equation}
Here, we have adopted the $\rm{eCMB+BAO+H_O}$ combined value of Table~4 in Ref.~\cite{Hinshaw:2012aka}.

In our case, lightest neutralino $\lspone$, i.e., the LSP, is the dark matter candidate. Instead of taking 
the $2\sigma$ window, we have respected only the upper limit of $\Omega h^2$. 
This is because the DM candidate need not to be necessarily single-component but can also be a multi-component 
one \cite{Berezhiani:1989fp,Boehm:2003ha,Ma:2006uv,Gu:2007gy,Hur:2007ur,Cao:2007fy,Hambye:2008bq,Aoki:2012ub,Heeck:2012bz}.
Although while presenting our benchmark points, we only consider those points which produce the perfect relic density, i.e, 
$0.1145 < \Omega h^2 < 0.1253$. 

\subsection*{II. Direct detection}

Apart from the relic density upper limit, we also discuss the implication of direct detection of DM using 
XENON100 \cite{Aprile:2012nq} and LUX \cite{Akerib:2013tjd} data on spin independent neutralino-proton $\lspone p$ scattering cross-section 
($\sigma_{\lspone p}^{SI}$). The $t$-channel Higgs and $s$-channel squark exchange diagrams contribute to $\sigma_{\lspone p}^{SI}$. 
For heavy squark scenario, the dominant contribution to the cross-section comes from the Higgs exchange diagram \cite{Drees:1993bu}. 
Again if $\lspone$ has sufficiently large higgsino component then $\sigma_{\lspone p}^{SI}$ may become large \cite{Hisano:2009xv}.

\section{Strategy of Analysis}
The prime aim of our analysis is to understand the correlations of the MSSM gaugino masses at the high scale itself which can explain the $[1\div 3]\sigma$ excess of muon (g-2). 
In the process, we have imposed several constraints, namely, bounds from collider searches and Higgs mass. We have also forced our solutions to respect  flavour constraints and the 
upper bound of  relic density. The recent searches for disappearing tracks and long lived charged particles by ATLAS and CMS have put very stringent constraints on the 
parameter space. In our analysis, particularly, these have played very crucial roles. 

To find the model independent correlations among $M_1,M_2$ and $M_3$ we have  treated these gaugino masses as individual free parameters and varied over  wide ranges along with other 
free parameters, like $m_0$ and $\tan\beta$. The tri-linear coupling $A_0$ is taken to be $-2m_0$ and Higgs parameters are set as $m_{H_u}^2=m_{H_d}^2=m_0^2$. Using these high 
scale input parameters, we have generated the SUSY spectrum at low scale and sorted out the parameter space which is compatible with the above mentioned constraints. We have noted the 
interconnections among different set of high and low scale parameters which have been discussed in detail in later section.  The same strategy has been implemented to identify 
the mixing  among  singlet and non-singlet contributions for different SUSY-GUT models from muon (g-2) window.

We have generated SUSY spectrum using SuSpect (v2.41) \cite{Djouadi:2006bz}. Further, for DM relic density, direct-indirect detection cross-section 
and flavour physics calculation we have used micrOMEGAs (v3.6.7) \cite{Belanger:2013oya} and calcHEP (v3.3.6) \cite{Belyaev:2012qa}. 
In our analysis, muon (g-2), has been computed using micrOMEGAs.

\section{High Scale Non-universality vs  Muon g-2}

\subsection{Non-universal gauginos and universal scalars - model independent analysis}\label{sec:Nonug_M-d_Ind}
For the sake of our analysis we have supplemented the following input parameters at the high scale: $m_0,M_1,M_2,M_3, \tan \beta, sgn(\mu)$. 
All the gaugino mass parameters ($M_1$, $M_2$, $M_3$) are varied randomly and individually  at the high scale over a wide range along with 
other parameters. Below we have listed  the ranges of parameters that we have considered for our detailed analysis: 
\begin{eqnarray}
m_0\in (1, 3000) ~{\rm GeV}; A_0 = -2m_0; \\ \nonumber
M_1\in (200, 5000) ~{\rm GeV}; M_2\in (-5000, +5000) ~{\rm GeV}; \\ \nonumber 
M_3\in (-5000, +5000) ~{\rm GeV}; \tan\beta\in (1, 60). 
\label{eq:scan_range}
\end{eqnarray} 

\begin{figure}[h!]
\centering
\includegraphics[scale=0.4,angle=270]{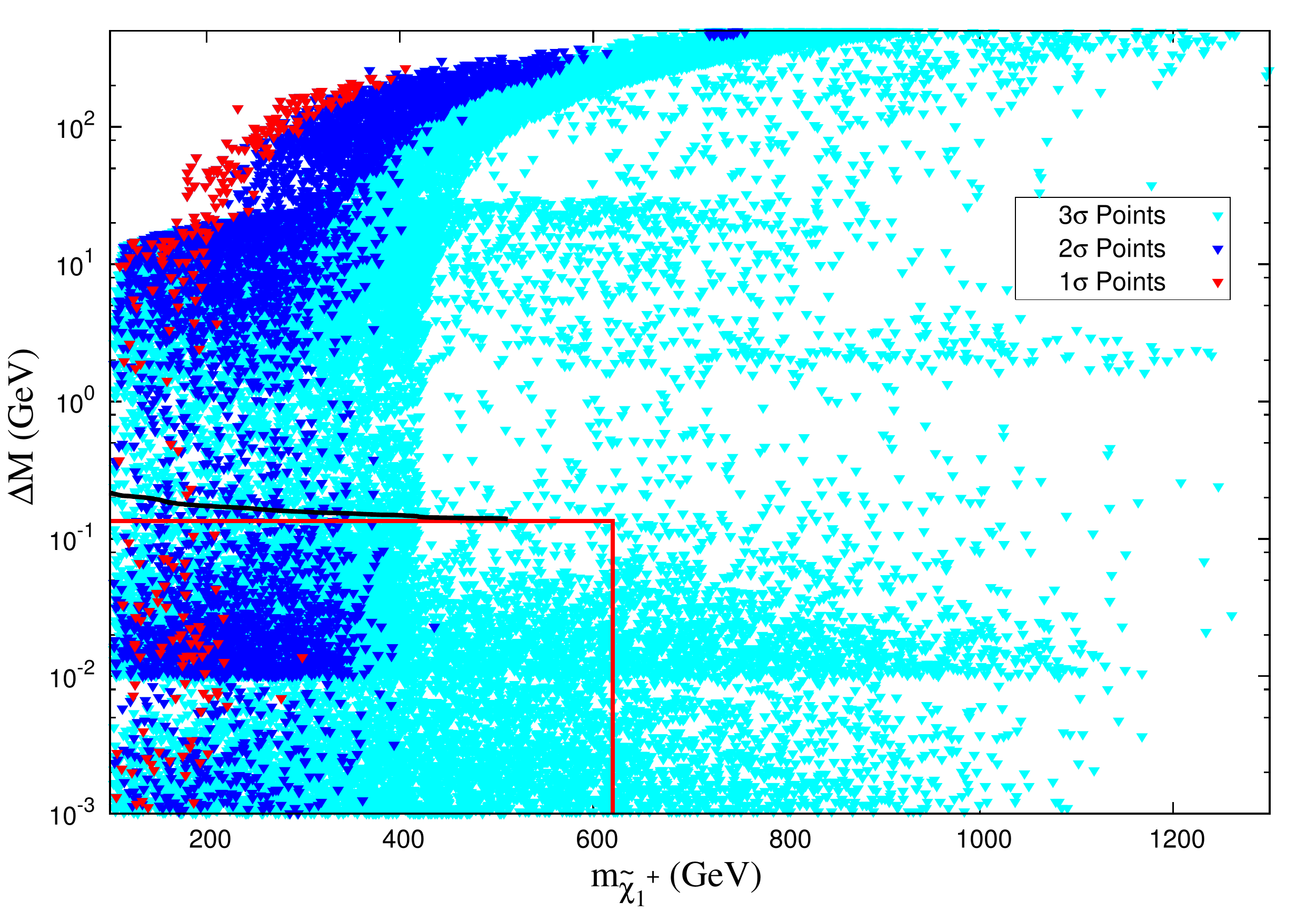}
\caption{Impact of constraints from search for disappearing track and long lived particles, 
see Sec.~\ref{sec:disappearing_track}. The region enclosed by the red lines is excluded from long-lived or stable charged particle search. 
The area under the slanted black line just above the horizontal red line is excluded from disappearing charged track search at the LHC.}
\label{fig:muong-2_char_excl} 
\end{figure}

We have mentioned earlier that the most stringent constraint is appearing for  nearly degenerate 
lighter chargino ($\mchonepm$) \& lightest neutralino ($\mlspone$) scenario.
We have noted that for a large part of the parameter space that leads to $[1\div 2]\sigma$ excess in 
$\Delta a_\mu$ contain nearly degenerate $\widetilde\chi^{\pm}_1$ 
and $\widetilde\chi^0_1$. Understandably, the $1\sigma$ points appear only at the low mass 
region for $\widetilde\chi^0_1$ 
and hence also for $\widetilde\chi^{\pm}_1$.  
In this type of scenario, a consequence of having a wino-like LSP is that the lighter chargino
can then be as light as the LSP, provided one has much heavier higgsinos what is exactly what we 
have here. If $\widetilde\chi^{\pm}_1$ mass is just above the $\widetilde\chi^0_1$ 
mass so that it can decay into a charged pion 
($\pi^{\pm}$) and the LSP, one observes a charge track at 
the detector. This puts a strong limit on the chargino mass \cite{ATLAS:2014fka}. Now if the degeneracy between the 
$\widetilde\chi^{\pm}_1$ and $\widetilde\chi^0_1$ masses is such that $\widetilde\chi^{\pm}_1$ cannot 
decay further, its mass limit becomes even stronger \cite{CMS:2014gxa,Aad:2013yna}. Fig.~\ref{fig:muong-2_char_excl} shows the distribution 
of the measure of  mass degeneracy $\Delta M$, as a function of  chargino mass.
The red, blue and cyan colours represent  the  points that signify  $1\sigma$, $2\sigma$ and $3\sigma$ 
excess in $\Delta a_{\mu}$  respectively. 
The exclusion limit \cite{CMS:2014gxa,Aad:2013yna} from long-lived or stable $\mchonepm$ is shown by the red rectangle at the left bottom corner of the plot. 
As one can see, all the points below $\Delta M = 135$ MeV
are excluded for 
a chargino mass upto 620 GeV. The region under the slanted black line just above the horizontal red line is excluded from disappearing track search of charged 
particle \cite{ATLAS:2014fka}. This, by far, 
appears to be the strongest constraint\footnote{In passing we would like to mention that the numbers we have quoted for the particle masses 
are obtained from SuSpect \cite{Djouadi:2006bz} which includes one-loop 
corrections \cite{Pierce:1994ew}. However, if one incorporate the two-loop corrections, the maximum mass splitting can be reduced by  2 to $5~{\rm MeV}$
\cite{Ibe:2012sx} depending upon the SUSY scale. Thus the real impact of this two loop correction 
will be hardly visible in our Fig.~\ref{fig:muong-2_char_excl}.} for a wino-like LSP scenario.

\begin{figure}[h!]
\centering
\includegraphics[scale=0.4,angle=-90]{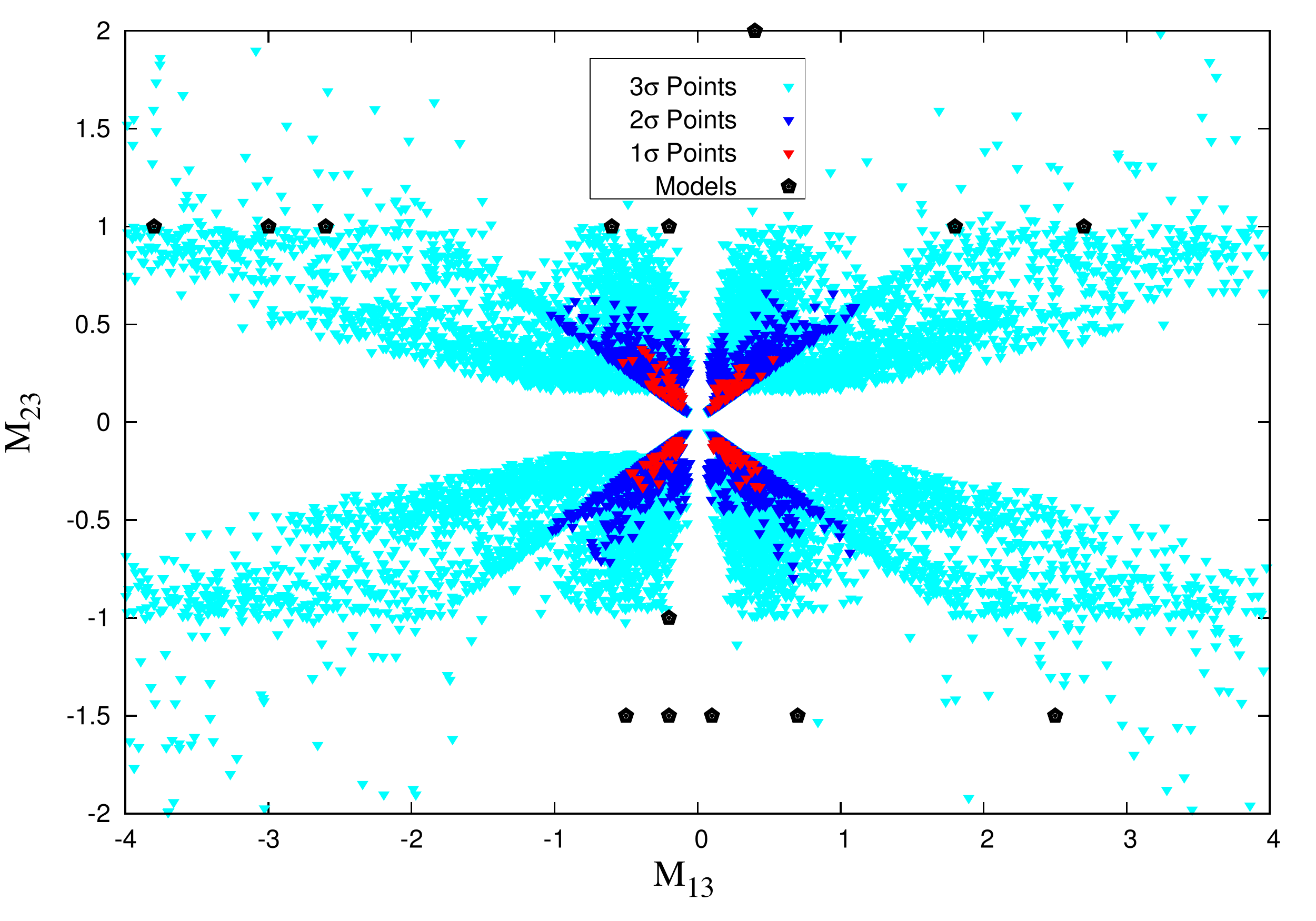}
\caption{\label{fig:muong-2_M13_M12} Model independent correlations among $M_1,M_2,M_3$ depending on $[1\div 3]\sigma$ muon (g-2) excess after satisfying all 
constraints discussed in the text. The black dots represent the models encapsulated in Tables 1, 2 in Ref.~\cite{Chakrabortty:2013voa}. Here, we have shown only few example model points where other models are out of this  frame.}
\end{figure}

From this point onwards, we consider only those points which survive the constraints imposed by the charged track and stable charged particle search
besides all the other collider, DM and flavour constraints as described in Sec.~\ref{constraints}. Colour coding corresponding to the 
$1\sigma$, $2\sigma$ and $3\sigma$ reaches of muon (g-2) remain unchanged in the rest of the paper.
In Fig.~\ref{fig:muong-2_M13_M12} we have shown the distribution of the obtained gaugino mass ratios ($\frac{M_i}{M_3}$, $i$=1,2).
We have chosen the range of gaugino masses  to encapsulate all the four possible correlations between the ratios $M_{13}$ ($\frac{M_1}{M_3}$) and $M_{23}$ 
($\frac{M_2}{M_3}$) including their signs.

Fig.~\ref{fig:muong-2_M13_M12} is almost symmetric and all the quadrants are in same footing for the purpose of our analysis.  
Hence it is sufficient to concentrate on one of the four quadrants to understand the nature of 
the distribution. It is clear that for an  enhancement of $\Delta a_\mu$ at $[1\div 2]\sigma$ level, 
one needs to have $|M_{23}| \leq 0.7$ and 
$|M_{13}|\leq 1.1$ excluding $M_{i3}=0$ for $i$=1,2. From these ratios, one can have an idea about the composition 
of the LSP under consideration. The mass splitting increases as the contribution of bino component increases in the LSP state. As evident from 
Fig.~\ref{fig:muong-2_char_excl} the mass splitting can be as large as $300~{\rm GeV}$ to produce the required 
$[1\div 2]\sigma$ enhancement in $\Delta a_{\mu}$. 

Note that, in this analysis, $\mu$ parameter is always relatively higher than $M_1$ and $M_2$ at the low scale. This a consequence of setting the Higgs soft masses ($m_{H_d}$ and 
$m_{H_u}$) equal to $m_0$ at the high scale.  Hence there is no possibility of having a small higgsino component that can contribute to the LSP state. 
Then it is evident that the most dominating contribution to the muon (g-2) enhancement would come from the processes involving light wino-like $\chonepm$
and light sleptons. This is one of the reason why the constraints from the search of heavy charged particles are so proactive in our case and rule out a 
large portion of the parameter space.          

\begin{figure}[h!]
\centering
\includegraphics[scale=0.29,angle=270]{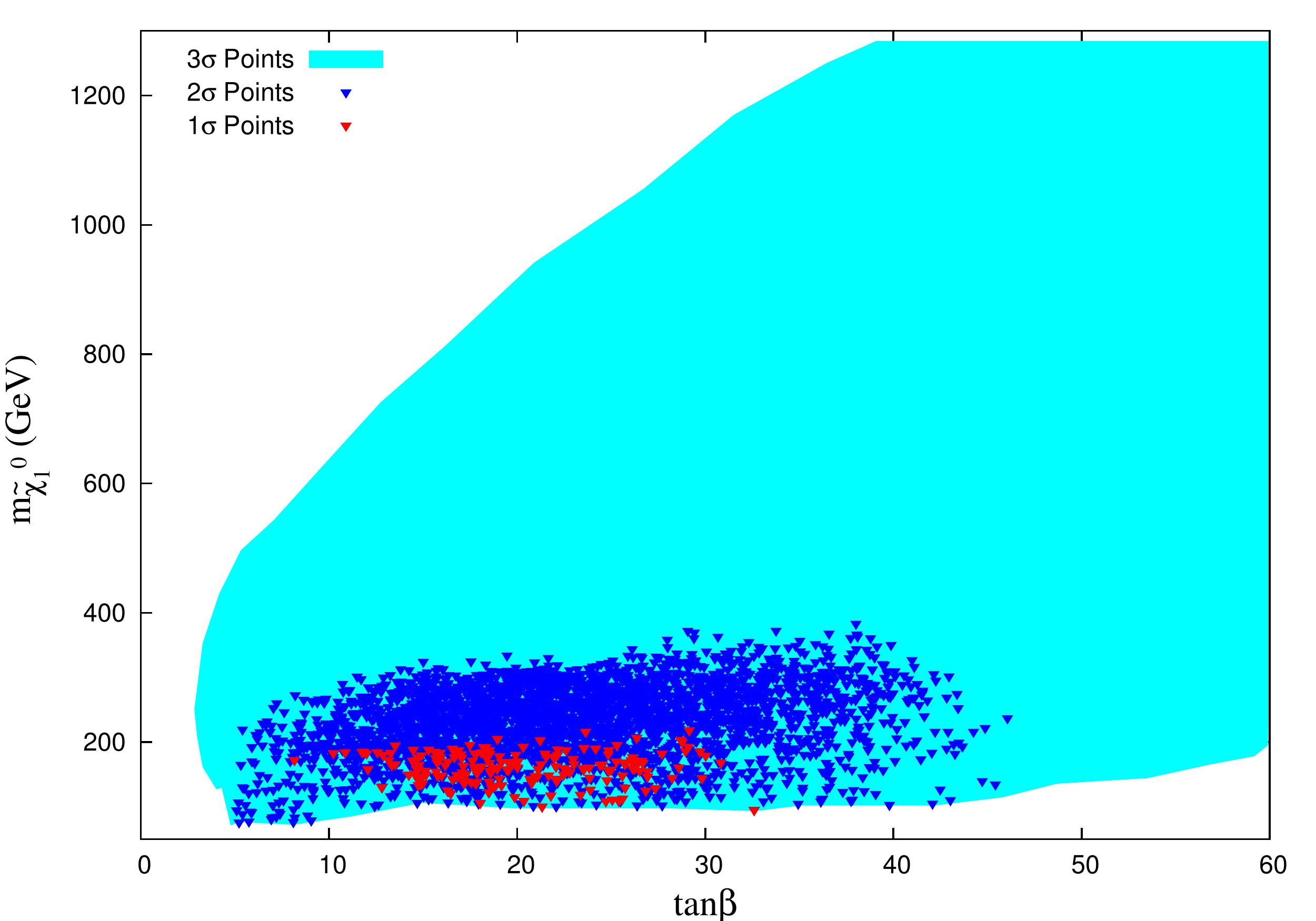}
\includegraphics[scale=0.29,angle=270]{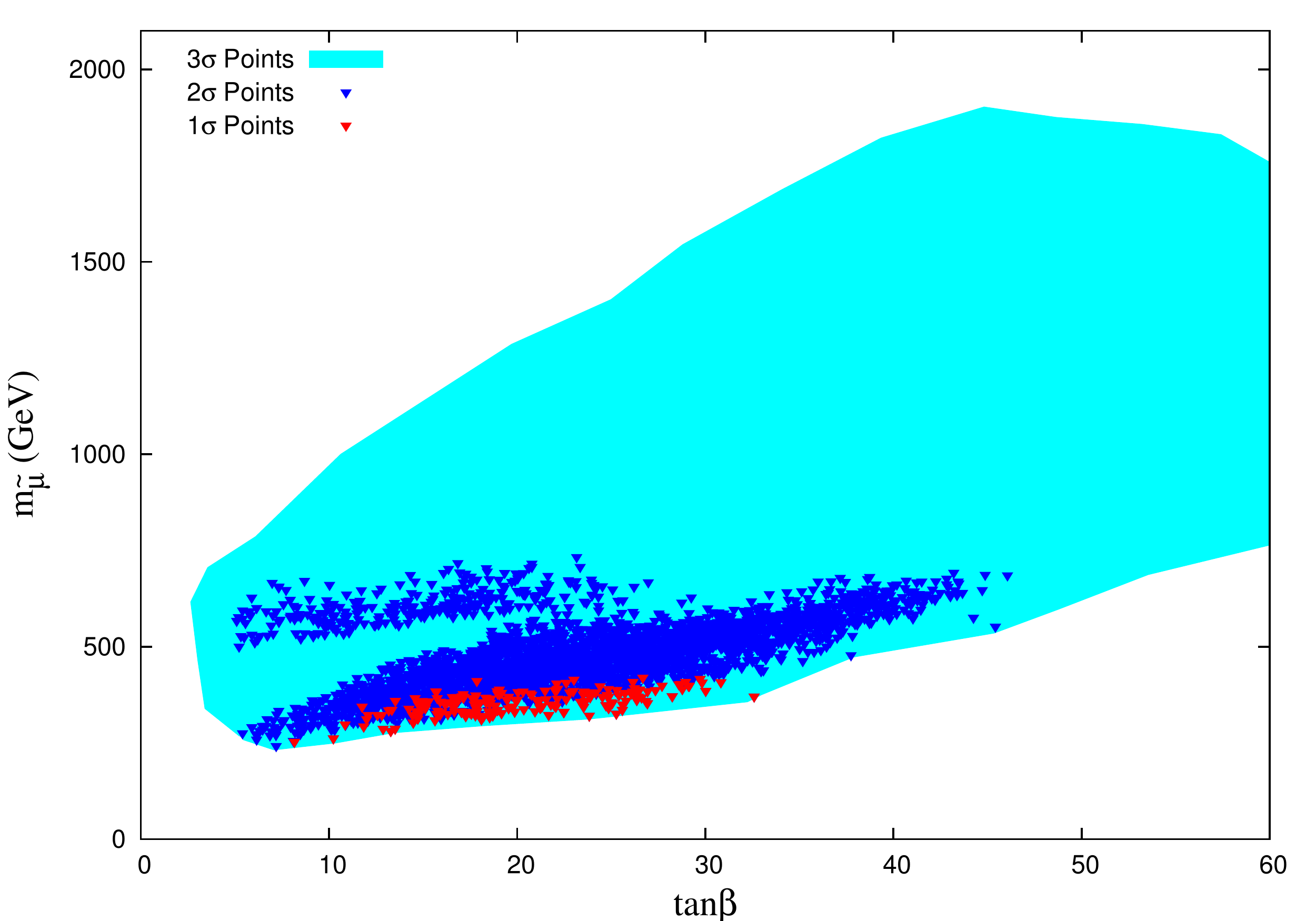}
\caption{\label{fig:muong-2_tanb} Correlations between low scale parameters: (Left panel) lightest neutralino mass 
($m_{\tilde{\chi}_1^0}$) vs $\tan \beta$; (Right panel) smuon mass 
($m_{\tilde{\mu}}$) vs $\tan \beta$, depending on $[1\div 3]\sigma$ muon (g-2) excess after satisfying all other 
constraints discussed in the text. A very similar correlation in $m_{\widetilde\nu}$ - 
$\tan\beta$ plane is found as given in  plot on the Right Panel. The slepton and sneutrino mass difference varies from $5$-$20$ GeV, indicating that the 
sneutrino-wino loop in Fig.~\ref{fig:muong-2_CSN} may also contribute significantly.}
\end{figure}

The other parameter that plays a very important role in the muon (g-2) calculation is $\tan\beta$. As discussed earlier,  
the SUSY contribution to muon (g-2) is directly proportional to $\tan\beta$. 
As a result, large $\tan\beta$, say $[8\div 35]$, is favoured to achieve $1\sigma$ enhancement which is evident from the Fig.~\ref{fig:muong-2_tanb}. 
However, the $2\sigma$ enhancement can be obtained for a $\tan\beta$ as low as 5 and as high as 47.

One can infer from the contributing diagrams in Fig.~\ref{fig:muong-2_NSM} that light slepton and LSP 
are crucial for the enhancement since they only appear as propagators. From Fig.~\ref{fig:muong-2_tanb}, 
we observe that the LSP mass must be lighter than 400 GeV for our purpose. Similarly, Fig.~\ref{fig:muong-2_tanb} 
puts an upper limit (700 GeV) on the smuon masses beyond which $2\sigma$ excess is not found. 

\begin{figure}[h!]
\centering
\includegraphics[scale=0.29,angle=270]{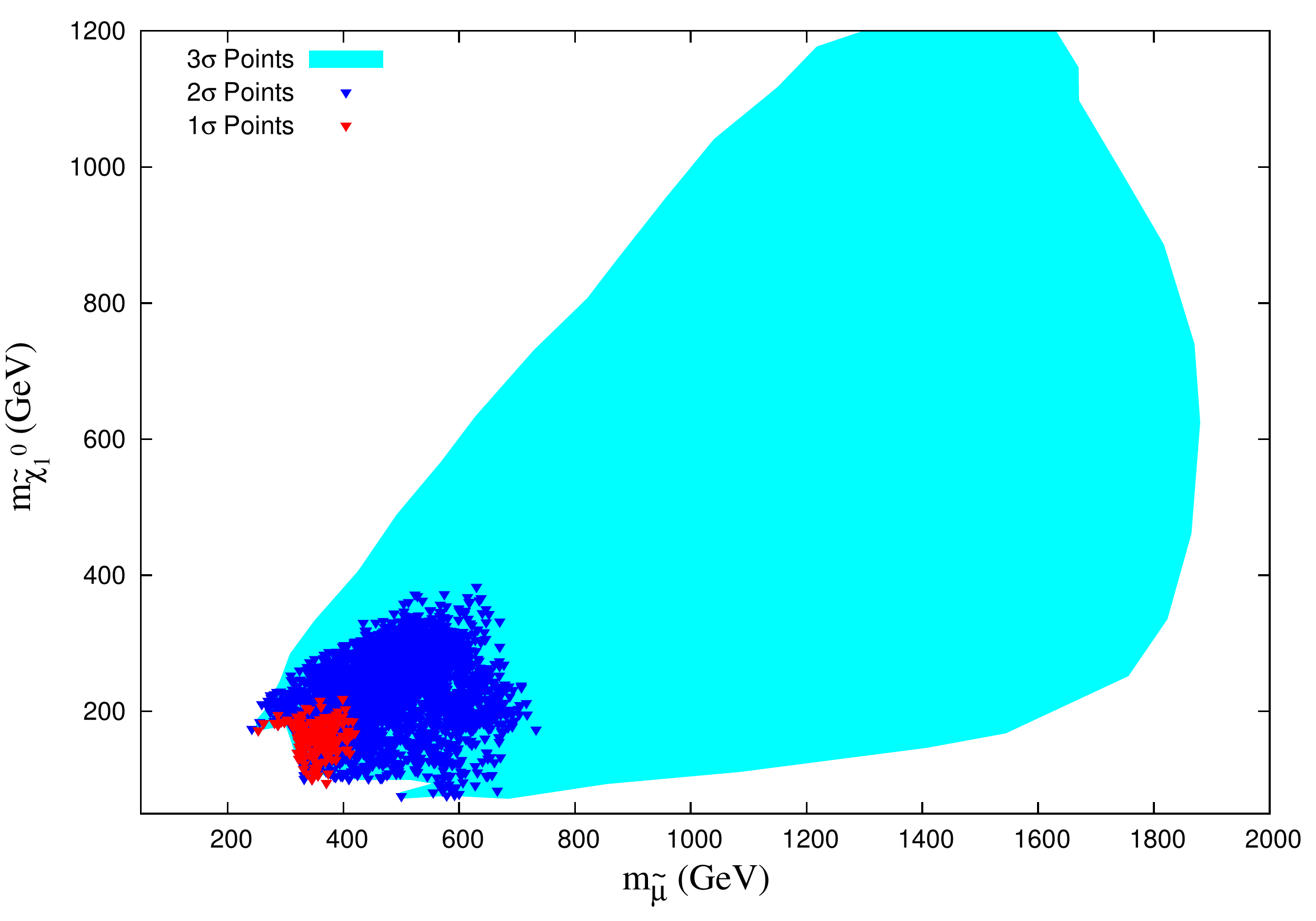}
\includegraphics[scale=0.29,angle=270]{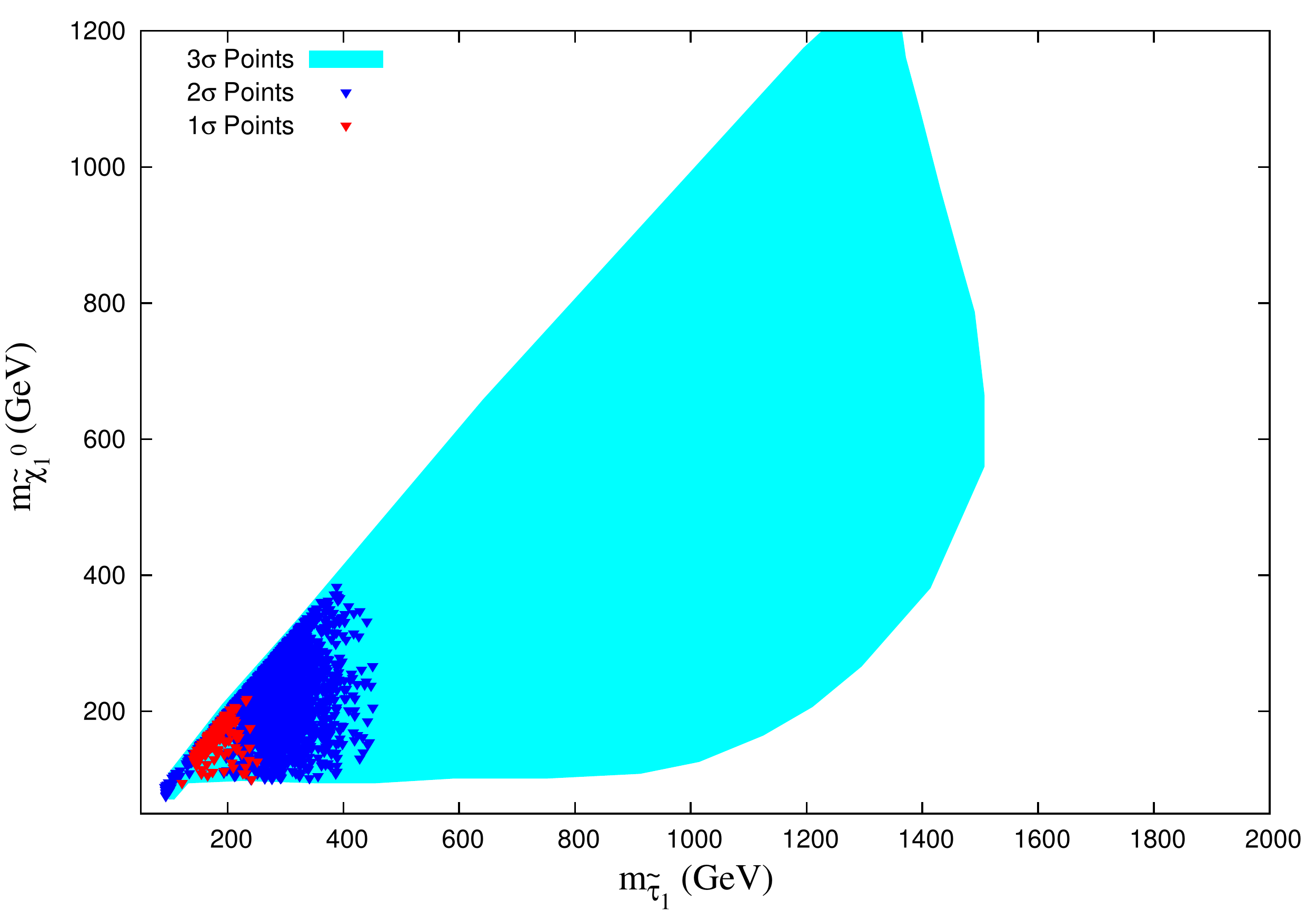}
\caption{\label{fig:muong-2_slep_lsp} Correlations between low scale parameters: (Left panel) lightest neutralino mass 
($m_{\widetilde{\chi}_1^0}$) vs smuon mass ($m_{\widetilde{\mu}}$); (Right panel) lightest neutralino mass  vs stau mass 
($m_{\widetilde{\tau}_1}$), 
depending on $[1\div 3]\sigma$ muon (g-2) excess after satisfying all other constraints discussed in the text.}
\end{figure}
Fig.~\ref{fig:muong-2_slep_lsp} shows the correlation between smuon 
mass and the LSP mass. Since the SUSY contribution to muon (g-2) decreases with the increase of  the smuon mass and(or) the LSP masses, 
we would expect a comet like structure in the smuon-LSP mass plane converging on the 
lighter side of both the masses, which is reflected in Fig.~\ref{fig:muong-2_slep_lsp} (Left). 
The plot shows only those points for which $\lspone$ is the LSP, which has been used as a constraint. Note that, a light stau($\widetilde\tau_1$) 
can also have an impact on the muon (g-2) calculation only if there is a sizeable mixing between smuon and stau states. 
For our choice of $A_0 (=-2m_0)$, we observe that this mixing may vary from negligible amount to as large as $10\%$ depending 
on the values of $m_0$, $M_3$ and $\tan\beta$. Hence we would expect a sizeable impact of the $\widetilde\tau_1$ mass in some part of the parameter 
space, specially, in the low slepton mass regime. Fig.~\ref{fig:muong-2_slep_lsp} (Right) shows the distribution of $m_{\tilde{\tau}_1}$ as a function 
of $m_{\tilde{\chi}_1^0}$.

\subsection*{Dark Matter Searches}

\subsubsection{Relic Density} 

\begin{figure}[h!]
\centering
\includegraphics[scale=0.29,angle=270]{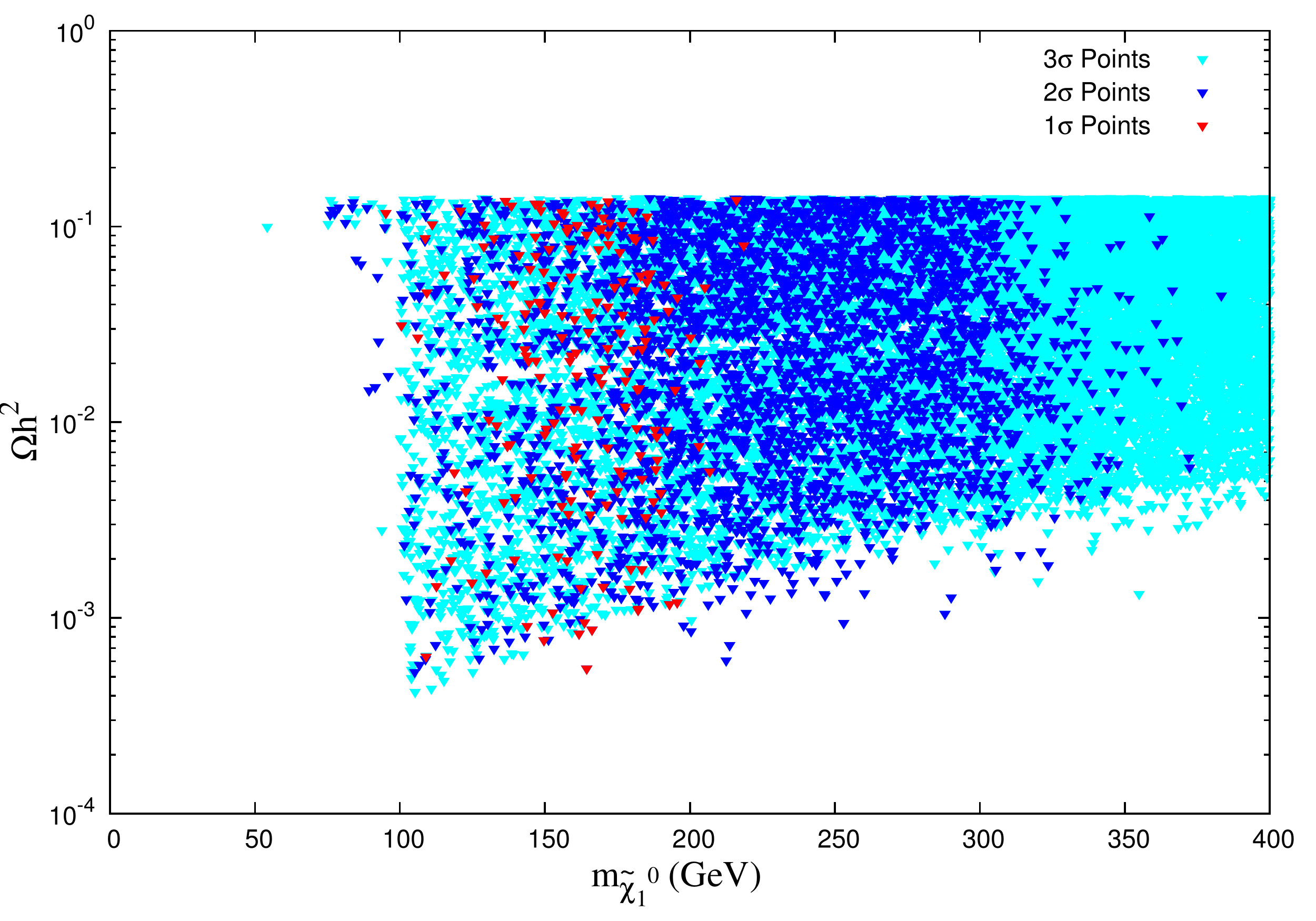}
\caption{\label{fig:relic_DM} Relic density as a function of the mass of the dark matter, i.e., LSP ($\tilde{\chi}_1^0$) 
depending on $[1\div 3]\sigma$ muon (g-2) excess after satisfying all other constraints discussed in the text.}
\end{figure}

In our analysis, the lightest neutralino $(\widetilde\chi^0_1)$ is the LSP and DM candidate since we have chosen to work in R-parity conserving scenario. 
We have checked the compatibility of muon (g-2) allowed parameter space with relic density constraint in Fig.~\ref{fig:relic_DM}. 
We consider only those points for which the DM annihilation and(or) co-annihilation are sufficient such that there is no over 
abundance of the lightest neutralino. 
As evident from Fig.~\ref{fig:relic_DM}, most of the favoured points produce a relic density that is much lower than the present upper limit. 
In our case, the LSP is composed  of  either bino or  wino or an admixture of both.

Dominant contribution to the DM relic density may come from 
bulk annihilation of $\widetilde\chi^0_1$ with sleptons (mostly $\widetilde\tau_1$), 
co-annihilation of $\widetilde\chi^0_1$ with the next to lightest supersymmetric particles (NLSPs) $\stauone$, $\chonepm$, $\lsptwo$. 
Thus the muon (g-2) and DM allowed parameter space can be classified into the following categories: 
\begin{itemize}
\item {\bf  Bulk annihilation:} The pure bulk annihilation region with light sleptons has been ruled out 
in usual mSUGRA scenario by the LHC constraints. The present exclusion in $m_0 - m_{1/2}$ plane \cite{Aad:2014wea,Aad:2014lra} is such 
that the sfermion is always much heavier than the LSP resulting in suppression of slepton mediated DM pair annihilation 
cross-sections. However, with  non-universal gaugino and universal scalar masses, 
we have noted that this can be a possibility and specially when $m_{\widetilde{\tau}_1} \le 130 ~{\rm GeV}$, it plays an important role 
in keeping $\widetilde\chi^0_1$ abundance under the specified limit. 
If other charged slepton masses are also  close to $m_{\widetilde\tau_1}$, they can participate in bulk annihilation as well 
with the final state consisting of two charged leptons. 
Benchmark points (BP) representing such scenario are illustrated in Table~\ref{tab:bp_nugm} (see BP1, BP2). 

\item {\bf stau co-annihilation:} 
In such scenario, $\widetilde\tau_1$ is the NLSP and the co-annihilation between $\widetilde\tau_1$ and LSP is significant to 
maintain the right relic abundance when $\lspone$ is mostly bino-like. Here the $\widetilde\chi^{\pm}_1$ and(or) $\widetilde\chi^0_2$ 
states are much heavier. 
For light $\widetilde\tau_1$ mass, dominant contribution comes from bulk annihilation process, 
whereas $\widetilde\tau_1$ - $\lspone$ co-annihilation contribution is roughly $\sim 15 -20 \%$ (see BP3 in Table~\ref{tab:bp_nugm}). 
With relatively larger $\mlspone$ (see BP4), the co-annihilation processes dominate. 

\item {\bf chargino co-annihilation:} 
Here, $\chonepm$ and(or) $\lsptwo$ appears as NLSP and consequently $\lspone$ annihilation with these sparticles are 
responsible to determine the relic density assuming the sleptons are much heavier to take part in co-annihilation. 
When the mass difference between LSP and NLSP is typically 15-25 GeV, one obtains PLANCK allowed DM relic abundance (see BP5 and BP6). 
The final states are usually dominated by quarks and gauge bosons. 
On the other hand, if $M_1$ is much larger than $M_2$, the $\widetilde\chi^{\pm}_1$ and $\lsptwo$ become nearly degenerate with the 
LSP and they co-annihilate profusely resulting in relic under-abundance. 

\item {\bf chargino and stau co-annihilation:} 
In such scenarios, $\widetilde\tau_1$, $\chonepm$ and(or) $\lsptwo$ masses are close to $\mlspone$. 
Along with bulk annihilation, stau co-annihilation and $\widetilde\chi^{\pm}_1$ and(or) $\lsptwo$  
co-annihilation give rise to correct relic density (see BP7 and BP8 in Table~\ref{tab:bp_nugm}).
\end{itemize}
\subsubsection{Direct and Indirect Detections}\label{sec:D-ID} 
\begin{figure}[h!]
\centering
\includegraphics[scale=0.29,angle=270]{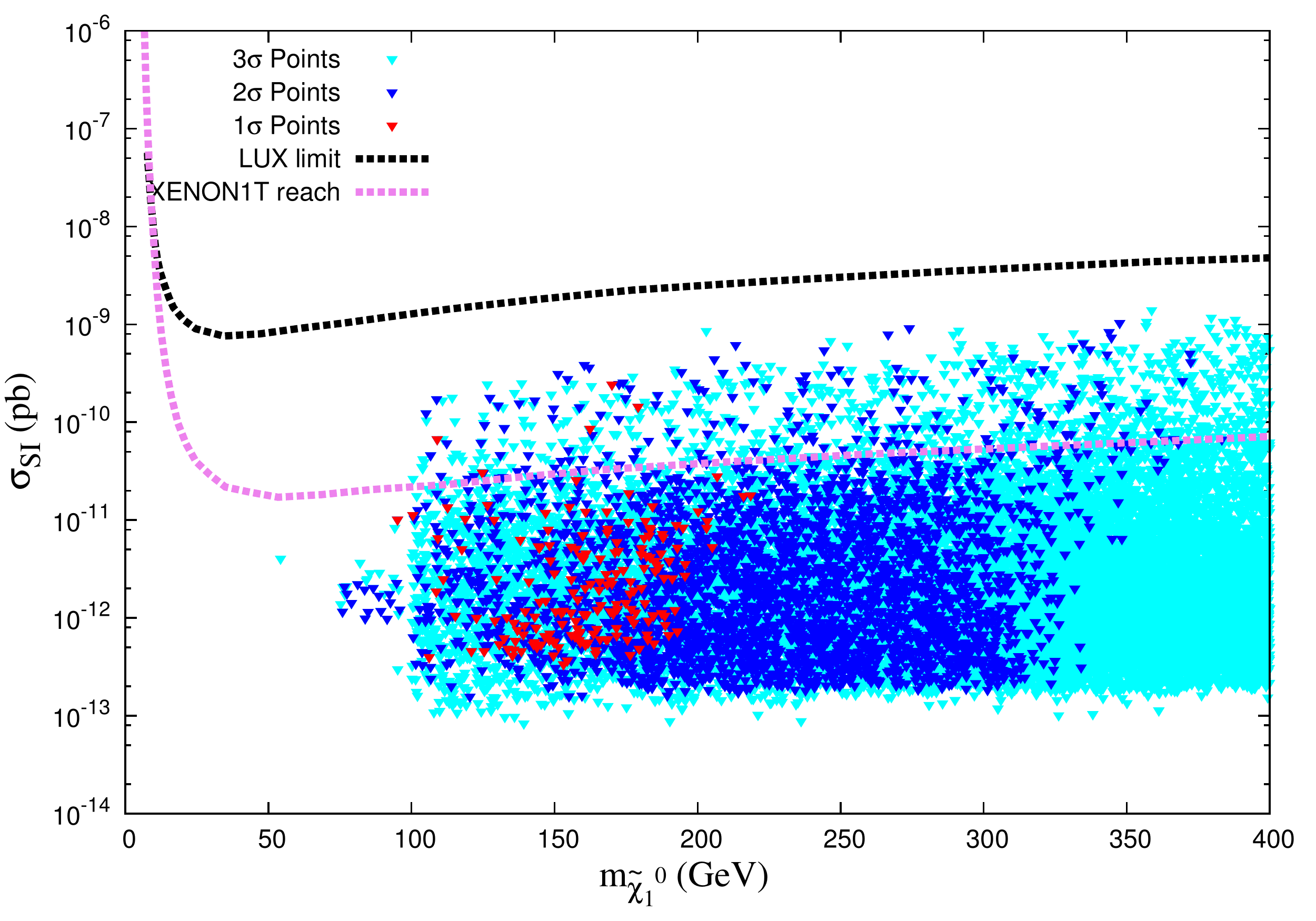}
\caption{\label{fig:sigsi_DM} Direct detection cross-section as a function of the mass of the dark matter, i.e., LSP ($\tilde{\chi}_1^0$) depending on $[1\div 3]\sigma$ muon (g-2) excess after satisfying all other constraints discussed in the text.}
\end{figure}

In Fig.~\ref{fig:sigsi_DM} we have shown the distribution of spin independent cross-section ($\sigma_{SI}$) as a function 
of the LSP mass. Note that, all the points shown in these plots obey the relic density upper limit.
As evident from the plot, apart from a small part of the $3\sigma$ allowed region, all the other points appear under the most recent exclusion line 
provided by LUX on $\sigma_{SI}$ \cite{Akerib:2013tjd} and thus allowed. 

The possible contribution to this cross-section comes from $t$-channel $Z$-boson, Higgs, squark mediated diagrams or 
$s$-channel squark mediated ones. The squarks being heavy do not contribute much. On the other hand, Higgs boson couplings to the first two generation quarks are 
suppressed. It is well known that within the MSSM framework, the DM-nucleon scattering cross-section may increase alarmingly if the LSP 
consists of a sizeable higgsino component which enhances its coupling with the $Z$-boson. Now in our scenario $\mu >> M_1, M_2$ at low scale for  most of the parameter space. 
Hence the LSP state always has a negligible higgsino component. In large LSP mass region, the $\mu$ parameter can be comparable with the other gaugino 
mass parameters and may enhance the scattering cross-section. However, this region of parameter space is not interesting from the $\Delta a_{\mu}$ enhancement viewpoint as discussed 
earlier. We have also shown in Fig.~\ref{fig:sigsi_DM}, the future projected limit of XENON1T \cite{Aprile:2012zx} experiment.
In case of a null result, it will reduce a significant portion of the available parameter space.
We have also taken into account the indirect detection cross-section constraints coming from 
different final states. All our points lie well within the exclusion limits and hence we do not present them in separate plots.

\subsubsection{Some Benchmark Points}
In this section, we have provided some benchmark points from different regime of the parameter space 
where $\Delta a_{\mu}$, $\Omega h^2$ and flavour constraints all lie within the $2\sigma$ ranges of their experimentally measured values. 
This parameter space is certainly compatible with  other collider and low energy constraints mentioned in earlier sections.

\begin{table}[h!]
\begin{tabular}{||c|c|c|c|c|c|c|c|c|c|c|c||} \hline\hline
Parameters  & BP1 & BP2 & BP3 & BP4 & BP5 & BP6 & BP7 & BP8 \\ 
\hline\hline 
$M_1$ & -241.8 & -248.5 & -330.1 & -647.7 & -274.4 & 516.2 & 373.5 & 581.4 \\
$M_2$ & 299.9 & -951.1 & -390.0 & 453.1 & 187.2 & -299.3 & 233.5 & 364.3 \\    
$M_3$ & 1004.1 & 4573.5 & 3075.0 & 1257.9 & 2275.3 & 857.1 & 1396.8 & 2636.8 \\
$m_0$ & 325.3 & 168.7 & 346.6 & 386.9 & 428.9 & 483.8 & 301.1 & 465.3 \\
$\tan\beta$ & 29.2 & 5.2 & 16.5 & 30.0 & 22.1 & 29.5 & 21.7 & 26.7 \\
\hline
$m_{\widetilde g}$ & 2205.5 & 9034.1 & 6269.1 & 2715.8 & 4751.0 & 1911.5 & 3003.5 & 5431.5 \\
$m_{\widetilde q_{L}}$ & 1937.9 & 7606.3 & 5327.8 & 2380.4 & 4074.1 & 1724.6 & 2600.7 & 4642.6 \\
$m_{\widetilde q_{R}}$ & 1938.6 & 7617.8 & 5345.4 & 2379.3 & 4089.8 & 1726.6 & 2609.4 & 4658.6 \\
$m_{\widetilde t_1}$ & 1582.8 & 6579.8 & 4594.1 & 1952.8 & 3479.0 & 1339.7 & 2189.8 & 3970.2 \\
$m_{\widetilde\ell_{L}}$ & 370.6 & 500.3 & 354.7 & 484.1 & 409.7 & 522.9 & 323.0 & 485.6 \\
$m_{\widetilde\ell_{R}}$ & 337.8 & 132.4 & 354.0 & 453.5 & 434.8 & 520.0 & 329.4 & 504.3 \\
$m_{\widetilde \tau_1}$ & 121.4 & 93.3 & 146.2 & 288.4 & 240.0 & 374.0 & 170.9 & 262.7 \\
$m_{\widetilde\nu_{1,2}}$ & 362.3 & 494.7 & 346.2 & 477.8 & 402.2 & 517.0 & 313.4 & 479.4 \\
$m_{\widetilde\nu_3}$ & 331.8 & 494.2 & 337.8 & 444.4 & 384.5 & 474.0 & 299.8 & 453.4 \\
$\widetilde\chi^0_1$ & 95.2 & 75.9 & 120.9 & 270.6 & 100.6 & 214.1 & 148.6 & 235.0 \\
$\widetilde\chi^0_2$ & 232.1 & 734.1 & 276.2 & 358.1 & 114.0 & 234.7 & 169.0 & 263.4 \\
$\widetilde\chi^{\pm}_1$ & 232.1 & 734.1 & 276.2 & 358.1 & 114.0 & 234.7 & 169.0 & 263.4 \\
$\mu$ & 1294.8 & 4863.9 & 3390.9 & 1582.2 & 2637.4 & 1198.1 & 1708.4 & 2990.0 \\
\hline
$(g-2)_{\mu}\times 10^{9}$ & 2.64 & 1.82 & 2.28 & 1.42 & 1.54 & 1.34 & 2.64 & 1.51 \\
\scriptsize{$\rm{BR}(b\rightarrow s\gamma)\times 10^{4}$} & 3.18 & 3.31 & 3.31 & 3.24 & 3.31 & 3.11 & 3.30 & 3.30 \\ 
\scriptsize{$\rm{BR}(B_s\rightarrow\mu\mu)\times 10^{9}$} & 3.81 & 3.07 & 3.07 & 3.38 & 3.08 & 3.74 & 3.12 & 3.09 \\
\hline
$\Omega h^2$ & 0.117 & 0.115 & 0.121 & 0.124 & 0.116 & 0.122 & 0.121 & 0.120 \\
$\sigma_{SI}\times 10^{13}$ (pb) & 101.0 & 11.6 & 4.6 & 87.1 & 6.3 & 281.9 & 38.6 & 6.73 \\
\hline\hline
\end{tabular}
\caption{High scale input parameters and the relevant sparticle masses along with the muon (g-2) value for some of the 
chosen benchmark points satisfying  the collider, DM and low energy constraints discussed in earlier sections. 
All the mass parameters are written in GeV unit.}
\label{tab:bp_nugm}
\end{table} 

In pure bulk (BP1 $\&$ BP2) or stau co-annihilation region (BP3 \& BP4), $m_{\widetilde\tau_1}$ is relatively light. Since we are working with 
universal scalar masses at the GUT scale, this implies either the slepton soft masses are light where the smuons 
are also expected to be light (see e.g. BP2) or $\tan \beta$ is effectively large. Both these scenarios may enhance $\Delta a_{\mu}$ 
which is reflected 
in BP1 - BP4. Light chargino may also produce the required enhancement provided the sneutrino masses are not too large. 
BP5 - BP6 represent chargino ao-annihilation region, where $\widetilde\chi^{\pm}_1$ is the NLSP. Hence the required enhancement in $\Delta a_{\mu}$ 
is obtained even with relatively larger smuon mass. BP7 and BP8 have almost degenerate $\widetilde\tau_1$ and $\widetilde\chi^{\pm}_1$ masses with large values of 
$\tan \beta$. Hence the combined effect of light chargino and large $\tan \beta$ produce the required enhancement in $\Delta a_{\mu}$.

A large tan$\beta$ results 
in light stau's in the spectrum. For such low scale spectra, from the production and subsequent decays of the gauginos, one would then 
expect $\tau$-enriched final states associated with missing energy. Since the $\tau$'s 
mostly decay hadronically, tagged $\tau$- jets + $\met$ signature will be the most suitable one to look for such scenarios. However, the LSP-NLSP mass 
gap being very small, the $\tau$-jet tagging efficiency will be very small. 

The squarks and gluinos are in general heavy in these spectra. Since the gluino mass is heavier than all the squark masses, it can 
decay into all the squark flavours abundantly giving rise to jets + $\met$ final state. In the usual mSUGRA scenario, on the other hand, 
usually the gluino decay via stop into multiple top and neutralinos resulting in the multi $b$-jet final state associated with large 
missing energy (see, for example, Ref. \cite{Aad:2014lra}). Moreover, given the present collider exclusion limits, in mSUGRA, with such heavy color sector as in 
our benchmark scenarios, it is not possible to achieve a slepton sector light enough to enhance the muon (g-2) value to the desired range.


\subsection{Universal gauginos and non-universal scalars - model independent analysis}
\label{sec:Ug_NUs_M_ind}
In this section, we have discussed the other possible high scale scenario in the present context, namely, universal gauginos with non-universal scalars. 

If we assume 
to have a universal mass parameter ($m_{1/2}$) for all the gauginos at the high scale, we expect the LSP to 
be mostly pure bino like at low scale\footnote{We are ignoring the higgsino possibility from the point of our analysis.}. 
This is due to the fact that the renormalisation group evolutions determine the gaugino mass ratios to be 
$M_1 : M_2 : M_3 \simeq 1 : 3 : 6$ at low scale. Hence in such scenario, the $\widetilde\chi^{\pm}_1$ contribution to the 
loop diagrams will be smaller. However, the bino component can be sufficiently light and can contribute to the enhancement, 
but then the relic density constraint can  be a serious problem as the 
DM pair annihilation cross-section may not be suitable  to satisfy proper relic abundance. The non-universal 
scalar scenario can be useful as it may allow  a stau ($\widetilde\tau_1$) 
to be light enough which can co-annihilate with $\widetilde\chi^0_1$ to make up for the annihilation 
cross-section. An added advantage of having a bino-LSP is that long-lived or stable 
chargino constraint which proved to be the most severe in the previous case, will not be applicable here.
 
In this scenario, the high scale input parameters are following: the universal gaugino mass ($m_{1/2}$), 
slepton mass ($m_0$) and squark mass ($m_0^{\prime}$). The soft Higgs mass parameters, $m_{H_d}$ and $m_{H_u}$ 
are assumed to be equal to $m_0$ at the GUT scale. The trilinear coupling, 
$A_0(= -2 m_0)$ is same as before. We note that this scenario can provide  $[2\div 3]\sigma$ 
excess in $\Delta a_\mu$ at its best. 
\begin{figure}[h!]
\centering
\includegraphics[scale=0.29,angle=270]{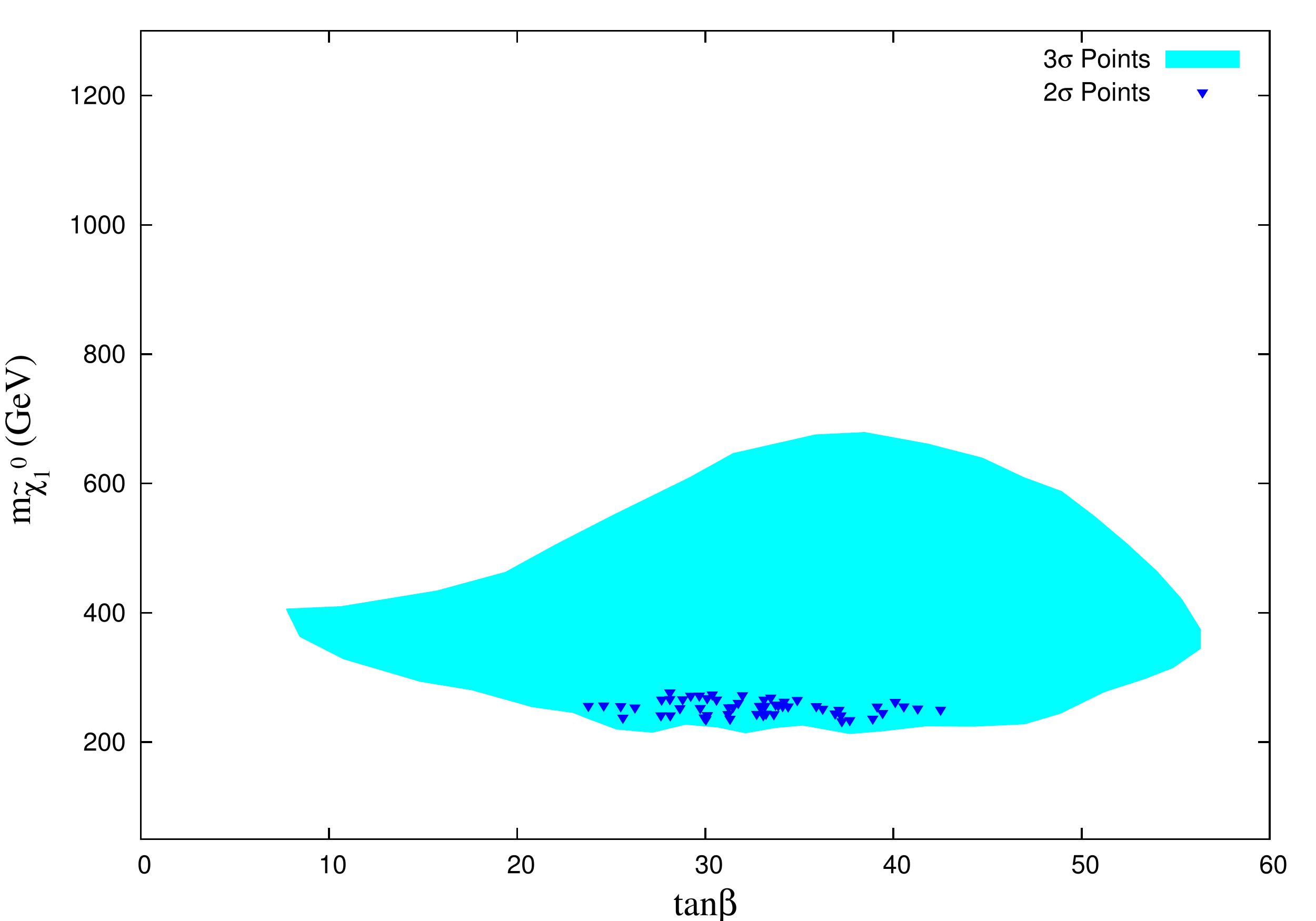}
\includegraphics[scale=0.29,angle=270]{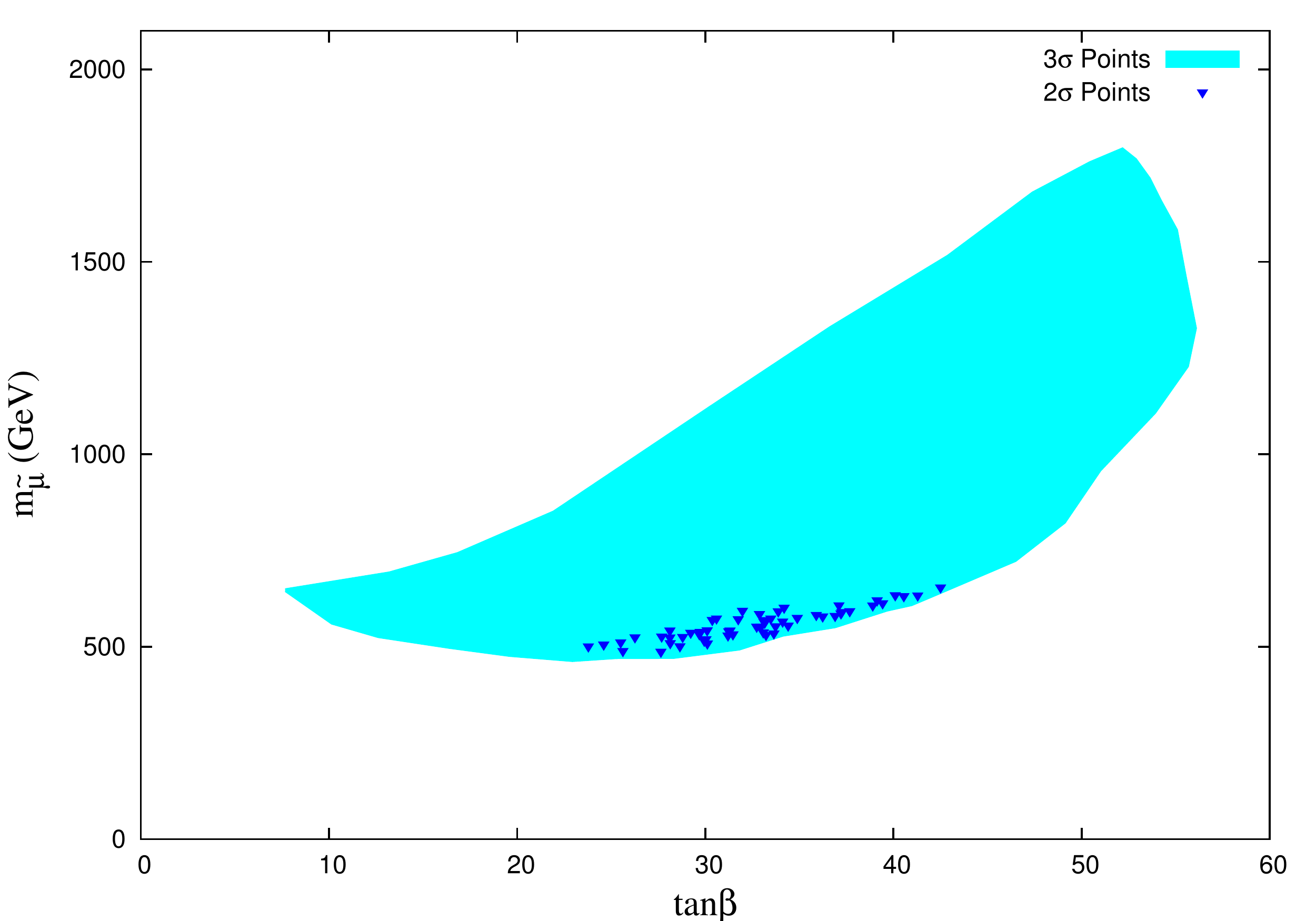}
\caption{\label{fig:muong-2:ugnus} Correlations between low scale parameters: (Left panel) lightest neutralino  mass 
($m_{\tilde{\chi}_1^0}$) vs $\tan \beta$; (Right panel) smuon mass 
($m_{\tilde{\mu}}$) vs $\tan \beta$, depending on $[2 \div 3]\sigma$ muon (g-2) excess after satisfying all other 
constraints discussed in the text. A very similar correlation in $m_{\widetilde\nu}$ - 
$\tan\beta$ plane is found as given in  plot on the Right Panel.  We have not found any parameter space in this scenario that  satisfy $1\sigma$ excess.}

\end{figure}

It is evident from Fig.~\ref{fig:muong-2:ugnus} that both the  slepton and LSP masses are heavier in this scenario
compared to the earlier case  at the low scale. The collider constraints for universal gaugino masses combine to produce these 
mass limits which prevent from achieving  $1\sigma$ excess of $\Delta a_{\mu}$. 
Assuming a universal gaugino mass at high scale the gluino appears to be roughly six times heavier than the LSP at low scale. 
Hence imposing a mass limit of $\sim 1.2~{\rm TeV}$ on the gluino mass for heavy squark scenarios automatically implies that the LSP cannot be lighter 
than $\sim 200~{\rm GeV}$. This scenario is being reflected in  Fig.~\ref{fig:muong-2:ugnus}. 

There can be another high scale scenario where both the gauginos and scalars are non-universal. For this case, 
the collider and DM constraints put very similar lower limits on the 
LSP and slepton masses to that we obtained for the non-universal gaugino and universal scalar scenario. The LSP 
composition is also very similar. Hence this scenario does not provide any new features to highlight. 


\subsection{Non-universal gauginos and universal scalars - model based analysis}

We have discussed the generation of gaugino masses in phenomenological $\n=1$ supergravity scenario in 
Sec.~\ref{mass_gen}. 
We have also noted that if the visible sector possesses  unified symmetry, then we need a non-singlet scalar 
to break that symmetry to achieve the SM gauge group. We have further discussed that in presence of a singlet 
and a non-singlet field in the hidden and visible sectors respectively, the generic gaugino mass terms can be 
written as $M_i  = M^{'} ~\left[1 +~\wp~ \delta_i~ \right]$ (see Eq.~\ref{eq:nonug_gaugino_mass_general}). 
This function $\wp$ is the ratio of $P,Q$ (see Eq.~\ref{eq:nonug_gaugino_mass}). The detail structure of $\wp$ is 
not easy to reveal. Thus we have encompassed the numerical ranges of $\wp$ which can explain the muon (g-2) 
excess successfully at $[1\div 2]\sigma$ level. The MSSM gaugino mass ratios at the high scale is depicted as:
\begin{eqnarray}\label{eq:Ratio_MSSM_gaugino}
M_1:M_2:M_3 = (1 +~\wp~\delta_1):(1 +~\wp~ \delta_2):(1 +~\wp~ \delta_3).
\end{eqnarray}

Earlier in Ref.~\cite{Chakrabortty:2013voa}, 25 different phenomenological models were analysed under 
the impression of dark matter search results. There contributions from the singlet scalar was neglected and 
prime focus was on the non-universal part only. We have revisited those models with their generic structures 
and from muon (g-2) point of view. Here, we have adjudged those non-universal models and checked which among them can successfully explain 
$[1\div 2]\sigma$  excess of $\Delta a_\mu$ (see Table~\ref{tab:M2ltM3}). We have analysed these models 
to find out the ranges of $\wp$ which is a measure of weighted mixing between singlet and non-singlet contributions.  
Note that among those 25 models (see Ref.~\cite{Chakrabortty:2013voa}), only 6 can successfully explain observed muon (g-2) excess at 
$[1\div 2]\sigma$ level after including the singlet contribution. Few models can provide $3\sigma$ excess but we are not quoting those here.
Note that, if this singlet contribution is neglected, all those 25 models fail to explain $\Delta a_{\mu}$ excess at $[1\div 2]\sigma$ level.
\begin{table}[h!]
\centering
\resizebox{\textwidth}{!}{
\begin{tabular}[c]{|c|c|c|}
\hline
Model Number &$M_1$ : $M_2$ : $M_3$  & Symmetry Breaking \\
 & (at M$_X$)   & Patterns\\
\hline\hline
1  &  -1/5 : -1 : 1  & $E(6)\xrightarrow{(35,1)\subset 650}(SU(6)\otimes SU(2)_R)$  \\
\hline
2 &  19/10 : 5/2 : 1   &  $SO(10)\xrightarrow{(1,1)\subset 770}(SU(4)\otimes SU(2)_R$   \\
\hline
3 &  1/10 : -3/2 : 1   & $E(6)\xrightarrow{(54,0)\subset 650}(SO(10)\otimes U(1))_{flipped}$   \\
\hline
4 &  -1/5 : -3/2 : 1   &  $E(6)\xrightarrow{(210,0)\subset 650,2430}(SO(10)\otimes U(1))_{flipped}$  \\
& & $(SO(10)\otimes U(1))_{flipped}\xrightarrow{(24)\subset 210}SU(5)$ \\
\hline
5 &  -1/2 : -3/2 : 1    &  $SO(10)\xrightarrow{(24)\subset 54,210,770} SU(5)$    \\
  &    & $SO(10)\xrightarrow{(24,0)\subset 54}(SU(5)\otimes U(1))_{flipped}$\\
  & &  $SO(10)\xrightarrow{(1,1)\subset 54}(SU(4)\otimes SU(2)_R)$\\
\hline
6 &  7/10 : -3/2 : 1     &  $SO(10)\xrightarrow{(24,0)\subset 210} (SU(5)\otimes U(1))_{flipped}$    \\
\hline
\end{tabular}
}
\caption{Gaugino mass models  that can explain $[1\div 3] \sigma$ excess of muon (g-2). Other models are not compatible with $[1\div 2]\sigma$ excess within this scheme.}
\label{tab:M2ltM3}
\end{table}
\begin{figure}[h!]
\centering
\includegraphics[scale=0.45,angle=270]{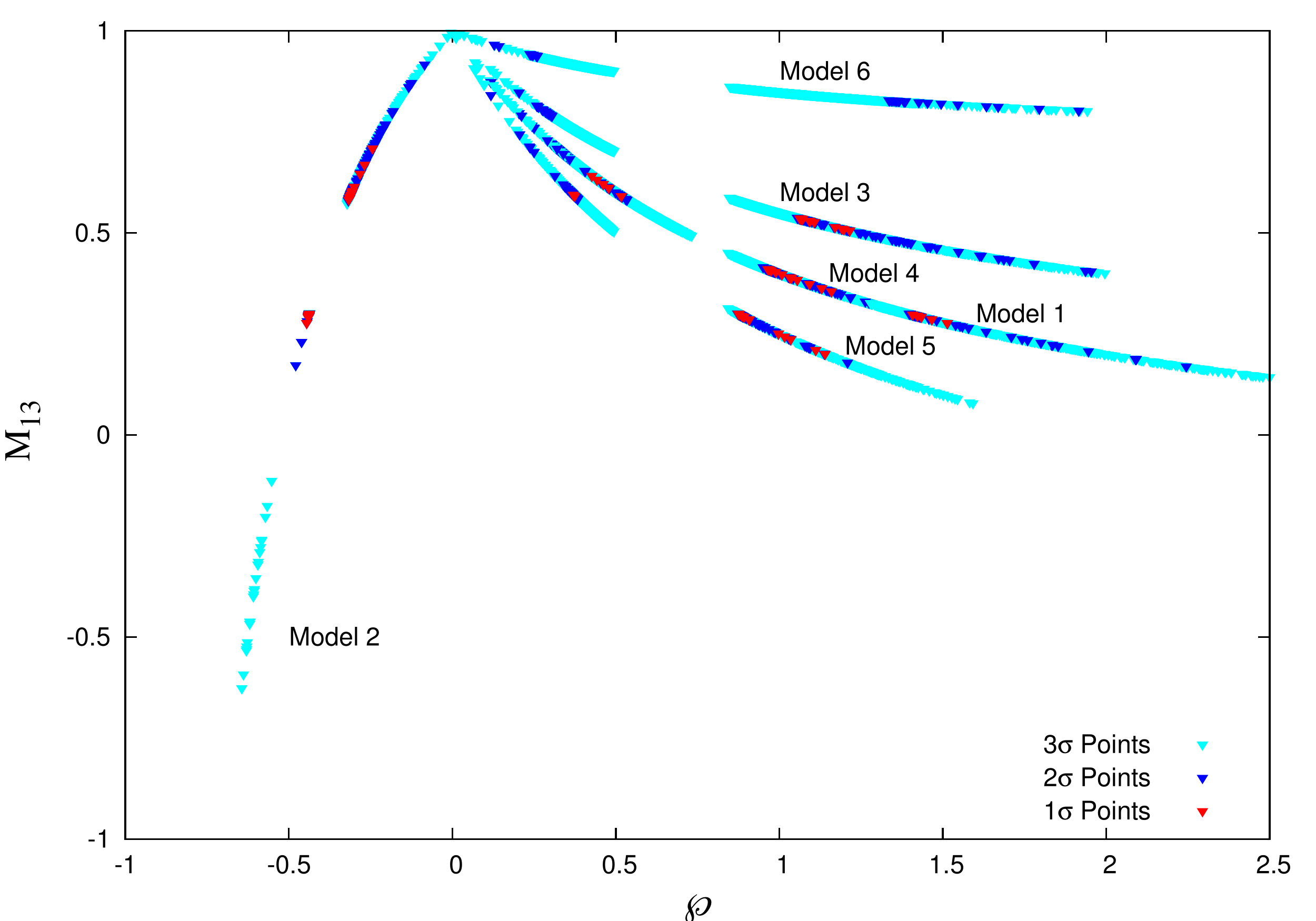}
\caption{\label{fig:muong-2_mixing_model2} The variations of mixing parameter  $\wp$ with $M_{13}$ for different set 
of models leading to $[1\div 3]\sigma$ excess in muon (g-2). The other models cannot provide $\Delta a_\mu$
excess at level of  $[1\div 2]\sigma$.}
\end{figure}

Fig.~\ref{fig:muong-2_mixing_model2} describes the correlation between $M_1/M_3$ and $\wp$ for different models. 
There are certain ranges of $\wp$ for which there exist  discontinuities in $M_{13}$ which are outcome of the fact 
that  within that range $M_{23}$ or $M_{13}$ vanishes. This happens because both are related for specific models. Around that 
solution, where $M_{23}$ or $M_{13}$ is very small, no parameter space is compatible with other imposed constraints. 
We have also explicitly shown the dependence of excess of muon (g-2) on the weighted mixing $\wp$.

\section{Conclusion}
High scale supersymmetry breaking is at the center of study since long time due to its minimalistic view. 
In this paper we have adopted a methodology to understand the correlations among high scale parameters in these 
type of scenarios, with a goal to explain the $[1\div 3]\sigma$ excess of  muon (g-2) over Standard Model predictions. 
To enrich the viability of our analysis, we have carefully incorporated the collider constraints without being biased 
to any particular scenario and also avoided the implementation of just mSUGRA bounds. The experimental constraints 
coming from simplified model assumptions have been tuned suitably. Hence the constraints 
imposed in our analysis can be taken over {\it mutatis mutandis}.

We have categorized our work in two parts: first we have discussed the model independent correlations among non-universal 
gaugino masses, and found out the moun (g-2) compatible solutions at $[1\div 3]\sigma$ level. We observe that none of 
the existing  models' predictions fit within this. Thus from muon (g-2) point of view, these models are not capable enough. 
We have  further analysed the high scale parameter space in terms of the low scale masses of sleptons, smuons, 
lighter chargino and lightest neutralino (LSP), along with $\tan \beta$.
Then we have also scrutinized such parameter space in the light of direct and indirect searches for DM. 
We have encompassed that regime by showing the dependence of  relic density and spin dependent cross-section ($\sigma_{SI}$) 
on mass of the DM.

We have briefly discussed the general structure of the gaugino masses that include the contributions from singlet 
and non-singlet chiral super-fields. Thus in such cases there is an extra parameter which is a measure of mixing of 
contributions from both fields. We have stuck to minimal models, i.e., restricted to one non-singlet field and 
explored the range of that mixing parameter from  muon (g-2) viewpoint along with other constraints. We have found 
only few of the existing  models can explain the excess in $\Delta a_{\mu}$ at $[1\div 2]\sigma$ level.  

In summary, we have captured the model independent features of non-universality considering muon (g-2) excess on a 
serious note. This broadly classified picture is also grabbed for some particular cases in terms of some specific 
benchmark points.


\section*{Acknowledgements}
Work of JC is supported by Department of Science \& Technology, Government of INDIA under the Grant Agreement number IFA12-PH-34 (INSPIRE Faculty Award). 
The work of AC, SM was partially supported by funding available 
from the Department of Atomic Energy, Government of India, for 
the Regional Centre for Accelerator-based Particle Physics 
(RECAPP), Harish-Chandra Research Institute.

\appendix 

\providecommand{\href}[2]{#2}
\addcontentsline{toc}{section}{References}
\bibliographystyle{JHEP}
\bibliography{nonug}
\end{document}